\documentclass[dvipsnames,journal]{IEEEtran}
    
\usepackage{subfiles}
\usepackage{algpseudocode}
\usepackage{algorithm}
\usepackage{cite}
\ifCLASSINFOpdf
   \usepackage[pdftex]{graphicx}
\else
   \usepackage[dvips]{graphicx}
\fi
\usepackage{color}
\usepackage{xcolor}

\ifCLASSOPTIONcompsoc
 \usepackage[caption=false,font=normalsize,labelfont=sf,textfont=sf]{subfig}
\else
 \usepackage[caption=false,font=footnotesize]{subfig}
\fi

\usepackage{bm}
\usepackage{bbm}
\usepackage{amsmath}
\usepackage{amsfonts}
\usepackage{cuted}

\usepackage{tabularx}
\usepackage{threeparttable} 
\usepackage{stfloats}
\usepackage{enumitem}
\usepackage[normalem]{ulem}

\usepackage{hyperref}
\hypersetup{
    colorlinks=true,
    linkcolor=black,
    filecolor=magenta,      
    urlcolor=teal,
    citecolor=black,
    pdftitle={Modal Statistics in Mode-Division-Multiplexed Systems using Mode Scramblers (Extended)},
    }
\usepackage[english]{babel}
\usepackage{amsthm}
\usepackage{amssymb}

\usepackage{xr}
\makeatletter

\newtheorem{theorem}{Theorem}
\newtheorem{corollary}{Corollary}[theorem]
\newtheorem{lemma}{Lemma}
\newtheorem{postulate}{Postulate}
\theoremstyle{definition}
\newtheorem*{definition}{Definition}
\theoremstyle{definition}
\newtheorem*{remark}{Remark}
\newenvironment{arguments}{{\noindent\textit{Arguments.}}}{\hfill$\square$}

\newcommand{\tr}[1]{\ensuremath{\mathrm{tr}\left({#1} \right)}}
\newcommand{\trspl}{\ensuremath{\mathrm{tr}}}
\newcommand{\E}[2][{}]{\ensuremath{\mathbb{E}_{#1}\left\{{#2} \right\}}}
\newcommand{\Espl}{\ensuremath{\mathbb{E}}}
\newcommand{\diag}[1]{\ensuremath{\mathrm{diag}\left\{{#1} \right\}}}
\newcommand{\PC}{\ensuremath{\bm{\mathcal{P}}}}
\newcommand{\DPC}{\ensuremath{\mathbf{D}^{-1} \bm{\mathcal{P}}}}
\newcommand{\tautau}{\ensuremath{\bm{{\tau}}}}
\newcommand{\gtot}[1]{\ensuremath{g_{\textrm{tot},{#1}}}}
\newcommand{\gtotv}{\ensuremath{\mathbf{g}_\textrm{tot}}}
\newcommand{\R}{\ensuremath{\mathbf{R}}}
\newcommand{\ttot}{\bm{\tau}_\textrm{tot}}
\newcommand{\bPhi}{\ensuremath{\bm{\Phi}}}
\newcommand{\bX}{\ensuremath{\mathcal{B}(\mathsf{X})}}
\newcommand{\prob}{\ensuremath{\textrm{prob}}}
\newcommand{\X}{\ensuremath{\mathsf{X}}}
\newcommand{\f}{\ensuremath{\mathbf{f}}}
\newcommand{\U}{\ensuremath{\mathcal{U}}}
\newcommand{\Ap}{\ensuremath{A_{+}}}

\newcommand{\RomanNumeralCaps}[1]
    {\MakeUppercase{\romannumeral #1}}

\newcommand{\betazs}{\tilde{\bm{\beta}}_0}
\newcommand{\betazsi}{\tilde{\bm{\beta}}_{0,i}}
    \newcommand\dBdwi{\tilde{\boldsymbol{\beta}}_{1,i}}
\newcommand{\tauS}{\ensuremath{\tilde{\bm{{\tau}}}}}
\newcommand{\tauSsub}{\ensuremath{\tilde{\bm{{\tau}}}_{\mathrm{\RomanNumeralCaps{2}}}}}

\newcommand{\tauSinit}{\ensuremath{\tilde{\bm{{\tau}}}_{\textrm{init}}}}
\newcommand{\tauSinitsub}{\ensuremath{\tilde{\bm{{\tau}}}_{\textrm{init},\mathrm{\RomanNumeralCaps{2}}}}}
\newcommand{\tauSZ}{\ensuremath{\tilde{\bm{{\tau}}}_{0}}}
\newcommand{\sigSZ}{\ensuremath{\tilde{{{\sigma}}}_{0}^2}}
\newcommand{\ddz}{\frac{\partial}{\partial z}}
\newcommand\dBdw{\tilde{\boldsymbol{\beta}}_1}

\newcommand\norm[1]{\left\lVert#1\right\rVert}
\newcommand{\Q}{\ensuremath{\tilde{\mathbf{Q}}}}
\newcommand{\DS}{\ensuremath{\tilde{\mathbf{D}}^{-1}}}
\newcommand{\DSinv}{\ensuremath{\tilde{\mathbf{D}}}}
\newcommand{\Qsub}{\ensuremath{\tilde{\mathbf{Q}}_{\mathrm{\RomanNumeralCaps{2}}}}}
\newcommand{\Qone}{\ensuremath{\tilde{\mathbf{Q}}_{\mathrm{\RomanNumeralCaps{1}}}}}

\newcommand\Qexp[1]{\exp{\left( -\left(#1\right)\Qsub\right)}}
\newcommand{\HH}{\ensuremath{\mathbf{H}}}

\newcommand{\RS}{\ensuremath{\tilde{\mathbf{R}}}}
\newcommand{\RSsubinit}{\ensuremath{\tilde{\mathbf{R}}_{\mathrm{\RomanNumeralCaps{2}}}}}

\newcommand{\RSsub}{\ensuremath{\tilde{\bm{\mathcal{P}}}}}

\newcommand{\PCS}{\ensuremath{\DS\tilde{\bm{\mathcal{P}}}}}

\newcommand\dBdwsub{\tilde{\boldsymbol{\beta}}_{1,\mathrm{\RomanNumeralCaps{2}}}}
\newcommand\dBinter{\tilde{\boldsymbol{\beta}}_{1,\textrm{inter}}}
\newcommand\dBintra{\tilde{\boldsymbol{\beta}}_{1,\textrm{intra}}}
\newcommand\dBO{\tilde{\boldsymbol{\beta}}_{1,\textrm{inter}}}
\newcommand\dBT{\tilde{\boldsymbol{\beta}}_{1,\textrm{intra}}}
\newcommand\lenT{L_{\textrm{intra}}}

\newcommand\lenS{L_S}
\newcommand\lenO{ L_{\textrm{inter}} }
\newcommand\lenTH{ \frac{L_{\textrm{mix}}}{2} }
\newcommand\lenOH{ \frac{L_{\textrm{inter}}}{2} }
\newcommand\DelBeta[1]{\Delta\beta_{1,#1}}
\newcommand{\vtil}{\tilde{\mathbf{v}}}

\newcommand\zone[2]{\zeta_1\left(#1,#2\right)}
\newcommand\ztwo[2]{\zeta_2\left(#1,#2\right)}

\DeclareMathOperator*{\minimize}{minimize}
\DeclareMathOperator*{\subjectto}{subject~to}

\begin{document}

\title{Modal Statistics in Mode-Division-Multiplexed Systems using Mode Scramblers (Extended)}%

\author{Anirudh~Vijay*,~\IEEEmembership{Student~Member,~IEEE,} Oleksiy~Krutko*,~\IEEEmembership{Student~Member,~IEEE,}\\Rebecca~Refaee and Joseph~M.~Kahn,~\IEEEmembership{Fellow,~IEEE}
\thanks{
* These authors contributed equally.
The authors are with the E.L Ginzton Laboratory, Department of Electrical Engineering, Stanford University, Stanford, CA 94305 (email: avijay@stanford.edu; oleksiyk@stanford.edu; becca24@stanford.edu; jmk@ee.stanford.edu).}%
\thanks{
\copyright~2024~IEEE. Personal use of this material is permitted. Permission from IEEE must be obtained for all other uses, in any current or future media, including reprinting/republishing this material for advertising or promotional purposes, creating new collective works, for resale or redistribution to servers or lists, or reuse of any copyrighted component of this work in other works.
}%
}%
\markboth{}
{Author \MakeLowercase{\textit{et al.}}: Insert Paper Title here}%

\maketitle%

\begin{abstract}
Typical multi-mode fibers exhibit strong intra-group mode coupling and weak inter-group mode coupling.
Mode scramblers can be inserted at periodic intervals to enhance inter-group coupling. 
The deterministic mode coupling of the mode scramblers, in concert with the random mode coupling of the fiber spans, can effect strong random mode coupling between all modes. 
This reduces both modal dispersion and mode-dependent loss, thereby decreasing receiver complexity and increasing link capacity.
In this paper, we analyze the effect of mode scramblers on end-to-end group-delay and mode-dependent loss standard deviations in long-haul multi-mode fiber links. We develop analytical tools in the generalized Jones and Stokes representations. We propose design criteria for mode scramblers that ensure strong end-to-end coupling: the mode-group-averaged power coupling matrix should be primitive and its non-dominant eigenvalues should be near zero.
We argue that when the mode scramblers satisfy these criteria, the probability distribution of the system transfer matrix asymptotically approaches that of a system with strong random mode coupling between all modes. Consequently, group-delay and mode-dependent loss standard deviations become sufficient statistics of the eigenvalues of the group-delay operator and the modal gains operator, respectively.
We also show that under certain conditions on the uncoupled group delays, it is possible to design self-compensating mode scramblers to reduce group delay accumulation below that of standard strong random coupling.
\end{abstract}

\IEEEpeerreviewmaketitle
\section{Introduction}

\IEEEPARstart{S}{pace-division} 
multiplexing (SDM) can increase capacity, integration, and power efficiency in long-haul optical coherent communication systems \cite{richardson_space-division_2013,essiambre_capacity_2012,puttnam_space-division_2021}.
SDM can be implemented using parallel single-mode fibers (SMFs), uncoupled-core or coupled-core multi-core fibers (MCFs), or multi-mode fibers (MMFs).
While each of these options has its pros and cons, mode-division multiplexing (MDM) in MMFs is considered attractive in achieving the highest level of integration \cite{winzer_chapter_2013}, for example, by enabling amplification using fewer pump modes than signal modes \cite{srinivas_efficient_2023}.

Long-haul MDM systems in MMFs need strong mode coupling to manage modal dispersion and mode-dependent gain and loss (collectively referred to as MDL) \cite{ho_linear_2014}.
Strong random coupling reduces the overall MDL standard deviation (STD), thereby increasing the average capacity and reducing the outage probability, and reduces the group-delay (GD) STD (also known as root-mean-squared GD spread), thereby reducing the digital signal processing (DSP) complexity at the receiver. Systems with strong random coupling exhibit GD and MDL STDs proportional to the square root of the length of propagation. 
Additionally, strong mode coupling improves frequency diversity, which further reduces the outage probability \cite{ho_frequency_2011,mello_impact_2020}.

Graded-index MMFs are often proposed for long-haul MDM links, owing to their relatively low uncoupled GD STD \cite{jensen_demonstration_2015,ryf_mode-multiplexed_2015}.
In these fibers, the spatial and polarization modes form mode groups.
The propagation constants of modes in the same mode group are nearly equal and those in different mode groups are significantly different.
Random perturbations in these fibers cause strong intra-group mode coupling and weak inter-group mode coupling \cite{fontaine_characterization_2013}.
One way of improving inter-group coupling is to periodically insert mode scramblers along the link \cite{arik_group_2016}.
Mode scrambling can be achieved in several ways, for example, using offset fiber launches \cite{raddatz_experimental_1998}, distributed multiple point-loads \cite{liu_wideband_2021}, long-period fiber Bragg gratings (LPFGs) \cite{zhao_broadband_2018,askarov_long-period_2015}, multi-plane light converters (MPLCs) \cite{li_design_2018}, {and using mode demultiplexer-multiplexer combinations} \cite{shibahara_long-haul_2020,di_sciullo_reduction_2023, hout_transmission_2024}.
Mode scramblers must have low intrinsic MDL STD, because signals in long-haul links may pass through hundreds of them. 
The mode scramblers with low MDL tend to be highly engineered and have deterministic transfer functions \cite{askarov_long-period_2015}. 
The analysis in this paper focuses on deterministic mode scramblers, such as those based on LPFGs.

Previous studies to design and analyze mode scramblers have employed heuristic objectives based on off-diagonal entries of the coupling matrix \cite{askarov_long-period_2015}. 
However, these criteria are restrictive, and evaluating mode scrambler performance requires brute-force multi-section simulations of the end-to-end system to quantify GD spread and overall MDL.
These multi-section simulations can significantly slow down iterative design procedures employing performance objectives of end-to-end GD or MDL STD.
There is a need for analytical tools that impose less restrictive design criteria and enable faster evaluation of performance objectives.

In this work, we study systems that interleave deterministic mode scramblers with GI-MMFs from first principles and develop analytical tools in the generalized Jones and the generalized Stokes representations.
We derive analytical expressions for end-to-end GD STD and MDL STD for such systems, pessimistically assuming no random inter-group coupling in the fiber.
We find that a mode-group power coupling matrix is a sufficient description of the mode scrambling induced by a mode scrambler. In particular, the structure and eigenvalues of the power coupling matrix emerge as the design criteria relevant for achieving strong coupling. 
We also provide a rigorous analysis to incorporate the effects of random inter-group coupling in the fiber and provide closed-form expressions for the GD STD.
We present simulations of systems with $6$ spatial modes ($12$ spatial and polarization modes) over $100$ spans of propagation, representative of long-haul and ultra-long-haul links. 
We argue that when the mode scramblers satisfy the aforementioned design criteria, the combination of deterministic mode scrambling and strong random intra-group coupling is asymptotically equivalent to strong random coupling between all modes. Consequently, the GD and MDL STDs become sufficient statistics of the group-delay operator and the modal gains operator, respectively, as the number of spans growns large.
We also show that the design space of mode scramblers includes a self-compensating regime wherein the GD STD periodically alternates between increasing and decreasing trends, similar to systems with group-delay compensating fibers.
Finally, our GD STD formulae are particularly useful in designing and optimizing mode scrambler devices, and can dramatically speed up evaluations of this objective in iterative design procedures. 


The remainder of the paper is organized as follows:
Section~\ref{sec:GD_MDL_MS} presents the study of GD and MDL STDs in the presence of mode scramblers in MDM-MMF systems using the generalized Jones representation under the assumption of strong intra-group coupling and no inter-group coupling in the fiber.
Subsections \ref{subsec:GDS} and \ref{subsec:MDL} provide analytical expressions for GD STD and MDL STD, respectively.
Subsection~\ref{subsec:MultiSection_sims} includes simulation results that verify the analytical expressions.
Section~\ref{sec:Design_criteria} provides design criteria for a strong mode scrambler.
Section~\ref{sec:Inter-Stokes} presents the study of GD STD in the presence of non-zero inter-group coupling in the MMF using the generalized Stokes representation. 
Section~\ref{sec:strong_random_coupling} presents the equivalence between mode-scrambler-aided coupling combined with strong random intra-group coupling and strong random coupling between all modes.
Section~\ref{sec:discussion} discusses some aspects of long-haul system design with mode scramblers.
Section~\ref{sec:conclusion} presents the conclusion.

{An extended version of this paper \cite{vijay_modal_2024} includes eight appendices providing derivations of key analytical results}. 

\section{GD Spread and MDL in the Presence of\\Mode Scramblers}
\label{sec:GD_MDL_MS}

\begin{figure}
\centering
\includegraphics[width=1\columnwidth]{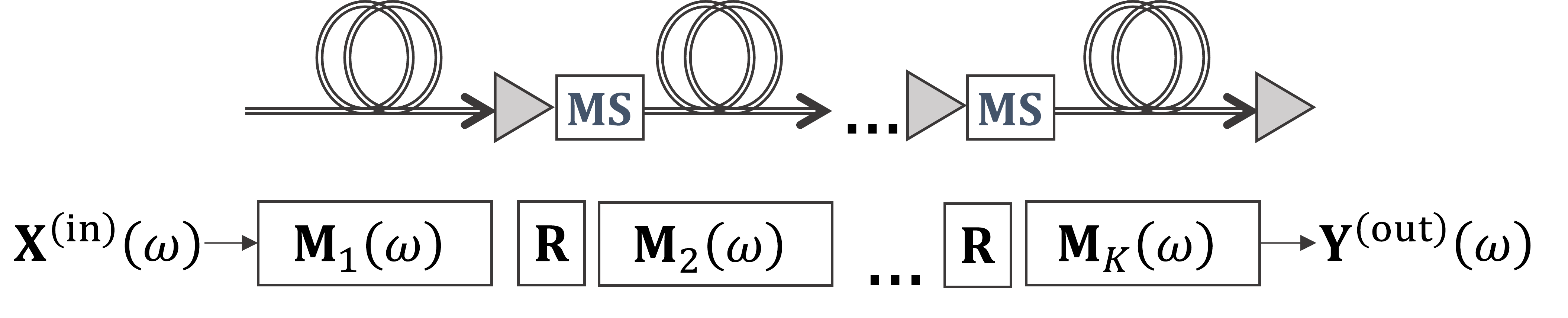}
\caption{Diagram of long-haul MDM transmission link over $K$ spans of propagation with periodic amplification and mode scrambling. MS: mode scrambler.}
\label{fig:sys_arch}
\end{figure}

A long-haul MDM transmission system using MMF with periodic amplification and mode scrambling is shown in Fig.~\ref{fig:sys_arch}. 
We define a \textit{span} as a section of MMF between successive mode scramblers. Each span is of length $\lenS$ and each span has identical statistical properties.
In the generalized Jones representation, $\mathbf{X}^{(\mathrm{in})}(\omega)$ and $\mathbf{Y}^{(\mathrm{out})}(\omega)$ denote the complex baseband input and output electric field vectors, respectively, in $D$ spatial and polarization modes at an angular frequency $\omega$. 
The matrices $\mathbf{M}_{1}(\omega),\mathbf{M}_{2}(\omega),\dots,\mathbf{M}_{K}(\omega)$  denote the $D\times D$ frequency-dependent random transfer matrices for the MMF spans including the amplifiers, and $\R$ denotes the $D\times D$ frequency-independent deterministic transfer matrix of each of the mode scramblers.
The $D$ modes are grouped into $N_g$ mode groups; modes in a mode group have nearly equal propagation constants. The set of modes in the $i$th mode group is denoted by $\mathcal{M}_i$ and the degeneracy is given by $d_i=\left | \mathcal{M}_i \right|$, where $\left | \cdot \right|$ represents the cardinality. 
The $D\times D$ transfer matrix of the system can be written as a product of the transfer matrices:
\begin{equation}
    \mathbf{M}_{\textrm{tot}}(\omega) = \mathbf{M}_{K}(\omega) \R \mathbf{M}_{K-1}(\omega) \R \dots \mathbf{M}_{2}(\omega)\R \mathbf{M}_{1}(\omega).
    \label{eq:Mtot}
\end{equation}

In the remainder of this paper, the $\omega$ dependence of the transfer matrices will not be explicitly shown.
For simplicity in analysis, we assume the mode-averaged gain of each amplifier perfectly compensates for the attenuation in the preceding MMF span, and any MDL caused by fiber splices, amplifiers, and mode scramblers is captured in $\R$.
Therefore, we can model $\mathbf{M}_{1}(\omega),\mathbf{M}_{2}(\omega),\dots,\mathbf{M}_{K}(\omega)$ by unitary matrices.

In typical GI-MMFs, strong intra-group coupling occurs over length scales of $\sim$1-500 m \cite{gabet_complete_2015,gruner-nielsen_measuring_2012}. During fiber manufacturing, processes such as spinning can also increase intra-group coupling strength.  On the other hand, inter-group coupling is weak and typically occurs over larger length scales. 
In published experiments, inter-group coupling has been observed over a wide range of lengths. Some report significant inter-group coupling after a few tens of kilometers \cite{jensen_demonstration_2015,ryf_mode-multiplexed_2015,genevaux_comparison_2014,bai_mode-division_2012}, whereas others report weak inter-group coupling even after $\sim 100$~km of propagation \cite{ryf_mode-division_2012}.

In this section, we assume strong intra-group and no inter-group coupling in the fiber to obtain conservative estimates of how the GD and MDL STDs depend on the mode scrambler transfer matrix $\R$. 
Therefore, the matrices $\mathbf{M}_{1}(\omega),\mathbf{M}_{2}(\omega),\dots,\mathbf{M}_{K}(\omega)$ can be modeled by block unitary matrices \cite{ho_linear_2014}.
The assumption of no inter-group coupling yields robust design criteria for mode scramblers in Section~\ref{sec:Design_criteria}.
Later, in Section~\ref{sec:Inter-Stokes}, we discuss how inter-group coupling in the MMF affects the GD STD.

\subsection{Analysis of GD Spread in the Generalized Jones Space}
\label{subsec:GDS}

In this subsection, we provide an analysis of GD STD for the system shown in Fig. \ref{fig:sys_arch} using the generalized Jones representation.
The group-delay operator (GDO) of the end-to-end link is defined as \cite{fan_principal_2005}:
\begin{equation}
\label{eq:gtot_def}
    \mathbf{G}_{\textrm{tot}}   ={} -\jmath \mathbf{M}_{\textrm{tot}}^{-1} \frac{\partial\mathbf{M}_{\textrm{tot}}}{\partial\omega},
\end{equation}
where $\jmath=\sqrt{-1}$. 
The GDs $\ttot$ are the eigenvalues of the GDO.
Since we are only interested in the GD STD, we assume that the eigenvalues sum to zero $\sum_{l=1}^D\tau_{\textrm{tot},l} = 0$.
The GD STD can be written as
\begin{align}
\begin{split}
    \sigma_{\mathrm{GD}}(K) ={}&\sqrt{\frac{\E{\norm{\ttot}^2 }}{D}} = \sqrt{\frac{\E{ \tr{\mathbf{G}^H_{\textrm{tot}} \mathbf{G}_{\textrm{tot}}} }}{D}},
\end{split}
\label{eq:gd_rms_tot}
\end{align}
where $\E{\cdot}$ denotes the expected value, $\norm{\cdot}$ represents the Eulidean or $l^2$-norm, $\tr{\cdot}$ denotes the matrix trace operation, and $\left(\cdot \right)^H$ denotes the conjugate transpose.

We can derive an analytical expression for $\E{\norm{\ttot}^2 }$ as a function of the number of spans $K$:
\begin{align}
    \begin{split}
        &\E{ \norm{\ttot}^2 }(K) \approx K\sum_{i=1}^{N_g} \sigma_{\text{intra},i}^2 + K \tautau_0^T \mathbf{D} \tautau_0 \\
    &+ \tautau_0^T \left(
    2 \sum_{k=1}^{K-1} \left(K-k \right) \mathbf{D}\left(\DPC \right)^{k} \right)\tautau_0,
    \end{split}
    \label{eq:gd_formula}
\end{align}
where $\tau_{0,i} = \lenS \sum_{l\in\mathcal{M}_i}\beta_{1,l}/d_i,~i=1,2,\dots N_g$ is the average GD of the $i$th mode group over a span of propagation, $\beta_{1,l}$ is the uncoupled GD per unit length of the $l$th mode such that $\sum_{l=1}^{D}\beta_{1,l}=0$, $\sigma_{\textrm{intra},i} \propto \sqrt{\lenS  L_{\textrm{intra},i}}$ is the intra-group GD STD of the $i$th mode group with intra-group coupling length $L_{\textrm{intra},i}$ over a span of propagation, $\mathbf{D}$ is the $N_g\times N_g$ diagonal matrix of mode group degeneracy, $\mathbf{D}[i,i] = d_i$, and $\PC$ is the $N_g\times N_g$ mode-group power coupling matrix of the mode scrambler. The $i,j$ element of $\PC$ is defined as the power transferred by the mode scrambler from mode group $j$ to mode group $i$:
\begin{equation}
\label{eq:PC_def}
    \PC[i,j] = \sum_{l\in\mathcal{M}_i}{\sum_{m \in\mathcal{M}_j}{ \left| \R[l,m] \right|^2}}.
\end{equation}
The diagonal elements of $\PC$ indicate the power retained by each mode group after transmission through the mode scrambler. In the case without mode scramblers, $\R$ and $\PC$ become diagonal matrices. 

The analytical expression in \eqref{eq:gd_formula} can be derived by expanding \eqref{eq:gtot_def} using the definition of $\mathbf{M}_{\textrm{tot}}$ in \eqref{eq:Mtot} and using results from random matrix theory. The derivation assumes MDL is negligible, in which case, $\R$ and $\mathbf{M}_\textrm{tot}$ are unitary. This approximation is valid when the overall MDL is low. 
A detailed derivation is provided in {\cite[Appendix B]{vijay_modal_2024}}.

The expression \eqref{eq:gd_formula} for GD STD indirectly depends on the mode scrambler transfer matrix $\R$ through the power coupling matrix $\PC$.
{Equation} \eqref{eq:gd_formula} simplifies the expression \eqref{eq:gd_rms_tot} containing $D\times D$-dimensional matrices to a sum containing real-valued scalars, $N_g$-dimensional vectors, and $N_g\times N_g$-dimensional matrices. 
The first two terms are proportional to $K$ and are independent of the mode scrambler. The first term is the intra-group coupling term. 
The second term is the sum of the squares of the average delays of the mode groups. 
The rest of the terms are dependent on the mode scrambler and form an arithmetico-geometric progression with ratio $\DPC$. 

{In the absence of a mode scrambler, $\R$ is a diagonal matrix and differs from the $D\times D$ identity matrix owing to MDL from the amplifier and from splicing. As a result, $\DPC$ is diagonal and close to the $N_g\times N_g$ identity matrix, and we get 
\begin{align*}
    \E{ \norm{\ttot}^2 }(K)     &\approx K\sum_{i=1}^{N_g} \sigma_{\text{intra},i}^2 + K \tautau_0^T \mathbf{D} \tautau_0 \\
                                &\qquad+ 2 \sum_{k=1}^{K-1} \left(K-k \right) \tautau_0^T\mathbf{D}\tautau_0 \\
                                &= K\sum_{i=1}^{N_g} \sigma_{\text{intra},i}^2 + K^2 \tautau_0^T \mathbf{D} \tautau_0,
\end{align*}
which is proportional to $K^2$ for large $K$. 
Consequently, $\sigma_{\mathrm{GD}}$ is proportional to $K$.
}
In the presence of a strong mode scrambler, we expect the sum of the arithmetico-geometric progression to be much smaller than the first two terms. Therefore, we can write $\E{ \norm{\ttot}^2 }(K) \approx K\sum_{i=1}^{N_g} \sigma_{\text{intra},i}^2 + K \tautau_0^T \mathbf{D} \tautau_0 $, which is proportional to $K$. Consequently, $\sigma_{\mathrm{GD}}$ is proportional to $\sqrt{K}$.


\subsection{Analysis of MDL in the Generalized Jones Space} 
\label{subsec:MDL}
In this subsection, we provide an analysis of MDL STD for the system shown in Fig. \ref{fig:sys_arch} using the generalized Jones representation.
The overall MDL of the system is described by the eigenvalues of the modal gain operator (MGO) defined as \cite{ho_mode-dependent_2011}:
\begin{equation}
\label{eq:Power_gains_operator}
    \mathbf{F}_\textrm{tot} =  \mathbf{M}_\textrm{tot} \mathbf{M}^H_\textrm{tot}.
\end{equation}
The MGO is Hermitian-symmetric and can be written as $\mathbf{F}_\textrm{tot}=\mathbf{V}\bm{\Lambda}^{(g)}_{\textrm{tot}} \mathbf{V}^H$, where $\bm{\Lambda^{(g)}}_{\textrm{tot}}=\diag{e^{\gtot{1}},\dots,e^{\gtot{D}}}$ is a diagonal matrix of positive eigenvalues representing the optical power gains and $\mathbf{V}$ is a unitary output beam-forming matrix \cite{ho_mode-dependent_2011}. Without loss of generality, the real-valued logarithmic power gains $\gtotv=[\gtot{1},\dots,\gtot{D}]^T$ sum to zero, $\gtot{1}+\dots+\gtot{D}=0$. 
The standard deviation of the overall MDL $\sigma_{\mathrm{MDL}}$ is a useful quantity for characterizing MDL \cite{ho_mode-dependent_2011} and is given by
\begin{equation}
\label{eq:MDL_overall}
    \sigma_{\mathrm{MDL}}(K) = \sqrt{\frac{1}{D}\E{\norm{\gtotv}^2}}.
\end{equation}
$\sigma_{\mathrm{MDL}}$ and $\gtotv$ are measured in log-power-gain units and can be converted to decibels by multiplying by $\gamma=10/\ln{10}\approx 4.34$, \textit{i.e.}, $\sigma_{\mathrm{MDL}}(\text{dB})=\gamma\sigma_{\mathrm{MDL}}(\text{log power gain})$. $\E{\norm{\gtotv}^2}$ can be written in terms of $\mathbf{M}_\textrm{tot}$ as 
\begin{equation}
    \label{eq:gtot2_general}
    \E{\norm{\gtotv}^2}(K) = \E{\tr{\left(\log \mathbf{M}_\textrm{tot} \mathbf{M}^H_\textrm{tot}\right)^2}}.
\end{equation}
In practical systems, wherein the overall MDL is low, the logarithm can be approximated by the first-order term. Under this approximation, we can derive an analytical expression for $\E{\norm{\gtotv}^2}(K)$:
\begin{equation}
\label{eq:gtot2_approx}
    \E{\norm{\gtotv}^2}(K)  \approx 2\left(\mathbf{d}^T  \left(\DPC\right)^{K-1}\mathbf{1}_{N_g}-D \right),
\end{equation}
where $\mathbf{d}$ is the column vector of mode-group degeneracies, and  $\mathbf{1}_{N_g}$ is the all-ones column vector of dimension $N_g$. 
Similar to the expression for GD STD, \eqref{eq:gtot2_approx} is also an $N_g\times N_g$-dimensional expression and depends on the mode scrambler transfer matrix $\R$ indirectly through the power coupling matrix $\PC$.
 A detailed derivation of \eqref{eq:gtot2_approx} is provided in {\cite[Appendix C]{vijay_modal_2024}}. 


\subsection{Multi-Section Simulations}
\label{subsec:MultiSection_sims}
\begin{table}
    \setlength\tabcolsep{0pt}
    \centering
    \caption{Simulation Parameters}
    \label{tab:siml_params}
    \begin{threeparttable}
        \begin{tabularx}{1\columnwidth} { >{\raggedright\arraybackslash}X  >{\centering\arraybackslash}X >{\raggedright\arraybackslash}X }
             \textbf{Parameter} & \textbf{Symbol}& \textbf{Value}  \\
             \hline \hline \\
             Number of spatial and polarization modes & $D$ & $12$\\
             Number of mode groups & $N_g$ & $3$\\
             Mode group degeneracy & $\mathbf{d}$ & $[2,4,6]^T$\\
             Number of spans  &$K$ & $100$\\
             \hline\\
             Uncoupled GDs per unit length& $\bm{\beta}_1$ & $[0.149, 0.149,$ $0.042, 0.042,$ $ 0.042, 0.042,$ $ -0.080, -0.080,$ $-0.076, -0.076,$ $ -0.076, -0.076]^T$ ps/km \\
             Span length &$\lenS$ & $50$ km\\
             Mode-averaged GDs over one span & $\tautau_0$ & $[25.2, 7.05,$ $-13.1]^T$ ns\\
             \hline \\
             Number of sections per span in the multi-section model & $N$  & $50$\\
             Number of fiber realizations &   & $10{,}000$\\
             \hline \\
             Per-span MDL STD & $\sigma_{\mathrm{MDL},0}$  & $0.2$ dB\\
             \hline \hline
        \end{tabularx}
            
    \end{threeparttable}
    
\end{table}

In this subsection, we verify the analytical expressions for GD and MDL STDs in Sections \ref{subsec:GDS} and \ref{subsec:MDL} through multi-section simulations similar to those discussed in \cite{ho_linear_2014,vijay_effect_2023}. 
The simulation parameters are provided in Table~\ref{tab:siml_params}. The mode indexing follows decreasing order of propagation constants: $[\{LP_{01,x},$ $ LP_{01,y}\},$ $\{ LP_{11a,x},$ $ LP_{11a,y},$ $ LP_{11b,x},$ $ LP_{11b,y}\},$ $\{ LP_{02,x},$ $ LP_{02,y},$ $ LP_{21a,x},$ $ LP_{21a,y},$ $ LP_{21b,x},$ $ LP_{21b,y}\}]$. 
Numerous fibers are generated using the parameters provided in Table~\ref{tab:siml_params}. Each of the $K$ spans is modeled as a concatenation of $N=50$ smaller sections with random block unitary coupling matrices to simulate strong intra-group coupling and no inter-group coupling. For $k=1,2,\dots,K$, we have
\begin{equation*}
    \mathbf{M}_k = \mathbf{T}_{k,N+1}\bm{\Lambda}_{k}\mathbf{T}_{k,N}\dots \mathbf{T}_{k,2}\bm{\Lambda}_{k}\mathbf{T}_{k,1},
\end{equation*}
where $\bm{\Lambda}_{k}$ is a frequency-dependent diagonal matrix given by 
$$\bm{\Lambda}_{k} = \mathrm{diag}\{e^{\left(\jmath \omega \beta_{1,1}\lenS/N\right)}, \dots,e^{\left(\jmath \omega \beta_{1,D}\lenS/N\right)}\},$$
and $\mathbf{T}_{k,1},\dots,\mathbf{T}_{k,N+1}$ are frequency-independent and independent and identically distributed (IID) random block unitary coupling matrices. Each of their block-diagonal components is independently distributed according to the Haar measure.

\begin{figure*}
\subfloat{\label{fig:PC_sims}}
\subfloat{\label{fig:gds_sims}}
\subfloat{\label{fig:mdl_sims}}
\centering
\includegraphics[width=\textwidth]{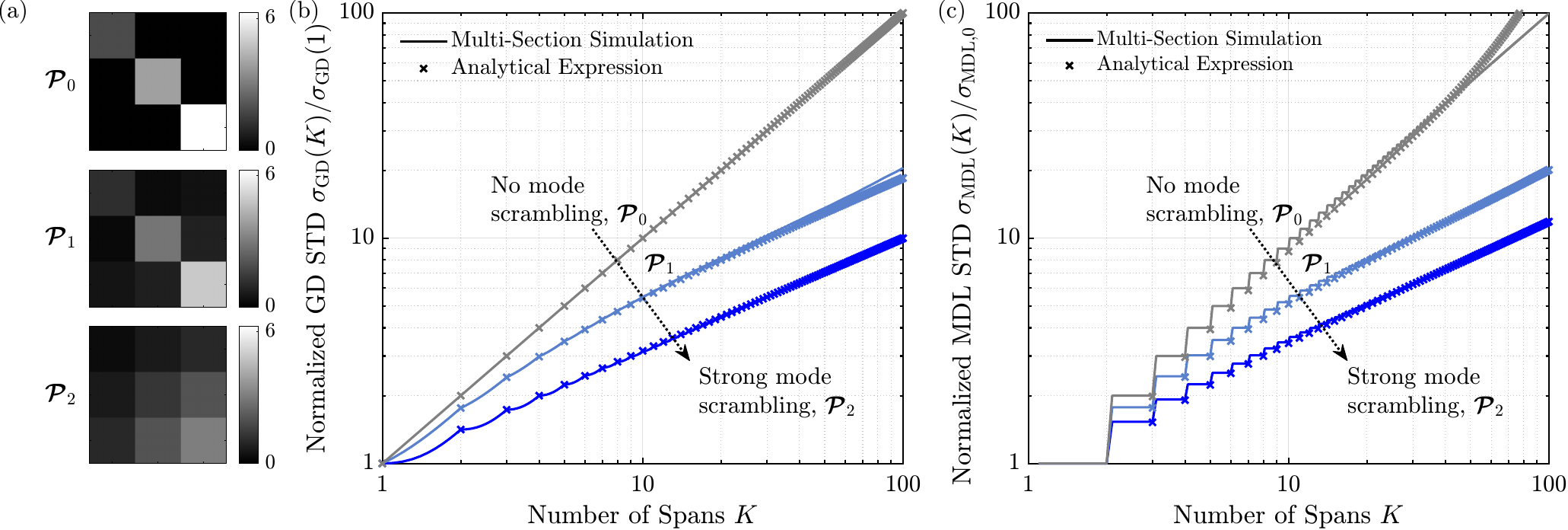}
\caption{Simulations of three mode scrambling scenarios. (a) Power coupling matrices; (b) GD STD as a function of the number of spans for three scrambling scenarios. The solid lines correspond to numerical estimates from multi-section simulations and the x-markers correspond to analytical estimates from the formula  \eqref{eq:gd_rms_tot}. The GD STD $\sigma_{\mathrm{GD}}(K)$ is normalized by the one-span GD STD $\sigma_{\mathrm{GD}}(1)$; (c) MDL STD as a function of the number of spans for three scrambling scenarios. The solid lines correspond to numerical estimates from multi-section simulations and the x-markers correspond to analytical estimates from the formula in  \eqref{eq:MDL_overall}. The MDL STD $\sigma_{\mathrm{MDL}}(K)$ is normalized by the per-span MDL STD $\sigma_{\mathrm{MDL},0}$. }
\label{fig:mdl_gds_sims}
\end{figure*}

To verify the GD and MDL STD formulae, we compare the GD spread and MDL performance in three scrambling scenarios.
\begin{enumerate}
    \item No mode scrambling: In this case, the mode scrambler transfer matrix is diagonal, $\R_0 =\bm{\Lambda}^{(g)}_0$, where $\bm{\Lambda}^{(g)}_{0}=\diag{e^{g_{0,1}/2},\dots,e^{g_{0,D}/2}}$ is the diagonal matrix of optical field gains, such that $g_{0,1}+\dots+g_{0,D} = 0$.
    \item Moderate mode scrambling: In this case, we generate a $D\times D$ unitary matrix $\bm{\mathcal{V}}$ that is almost diagonal and incorporate MDL, $\R_1 =\bm{\mathcal{V}}^{0.5}\bm{\Lambda}^{(g)}_0\bm{\mathcal{V}}^{0.5}$.
    \item Strong mode scrambling: Intuitively, any mode scrambler with a dense transfer matrix is a suitable candidate. Thus, we pick the normalized DFT matrix\footnote{The $l,m$ element of the DFT matrix is $\mathbf{W}_D[l,m]=(1/\sqrt{D})*\exp{(2\pi \jmath(l-1)(m-1)/D})$.} 
    $\mathbf{W}_D$ of size $D\times D$ and incorporate MDL, $\R_2 =\mathbf{W}_D^{0.5}\bm{\Lambda}^{(g)}_0\mathbf{W}_D^{0.5}$.
\end{enumerate}
{In all three cases, $\mathbf{g}_0$ is a real-valued vector chosen such that $g_{0,1}+\dots+g_{0,D}=0$ and the per-span MDL STD,
$\gamma\sigma_{\mathrm{MDL},0} = \gamma\sqrt{\tr{\left(\log \R \R^H\right)^2} / D}=0.2\,\text{dB}$.
It is to be noted that the sources of MDL, namely, mode scrambler devices, amplifiers, and fiber slices, are assumed to be lumped at the end of a span. Amplifier and fiber splice transfer matrices can be modeled by appropriate non-unitary block-diagonal matrices. Owing to the assumption of strong intra-group coupling, this modeling is redundant. We have verified through simulations that in the absence of mode scramblers, the GD and MDL STDs caused by such devices are the same as those obtained with a diagonal transfer matrix $\R_0$ when the per-span MDL STD is scaled appropriately.}
The power coupling matrices $\PC_0$, $\PC_1$ and $\PC_2$ corresponding to $\R_0$, $\R_1$ and $\R_2$, respectively, are shown in Fig.~\ref{fig:PC_sims}.

For the three scrambling scenarios, the GDO in \eqref{eq:gtot_def} and the MGO in \eqref{eq:Power_gains_operator} are computed after every fiber section and for each of the fiber realizations. The eigenvalues are then extracted to obtain numerical estimates of the GD and MDL STDs. Analytical estimates of GD STD \eqref{eq:gd_rms_tot} and MDL STD \eqref{eq:MDL_overall} are computed after every span of propagation.

Fig.~\ref{fig:gds_sims} shows the normalized GD STD $\sigma_{\mathrm{GD}}(K)/\sigma_{\mathrm{GD}}(1)$ plotted as a function of the number of spans $K$. As indicated by the slope of the log-log plot, in the case of no mode scrambling, the GD STD increases linearly with the number of spans. 
In the case of strong mode scrambling, the GD STD increases with the square root of the number of spans. These observations are consistent with the behavior of GD STD in weak and strong random coupling regimes, respectively. 
In all cases with low overall MDL, the estimates from the analytical expression in \eqref{eq:gd_rms_tot} closely match the numerical estimates from the simulation. The discrepancy between the estimates for large $K$ in the moderately scrambled scenario can be attributed to a significant overall MDL, so the assumption that $\mathbf{M}_\textrm{tot}$ is unitary does not hold.


Fig.~\ref{fig:mdl_sims} shows the normalized MDL STD $\sigma_{\mathrm{MDL}}(K)/\sigma_{\mathrm{MDL},0}$ plotted as a function of the number of spans $K$. $\sigma_{\mathrm{MDL},0}~=~\sqrt{\tr{\left(\log{\R \R^H}\right)^2}/D}$ is the per-span MDL STD.
The step-like pattern of the multi-section simulations results from the MDL sources being localized at the ends of the spans, unlike modal dispersion, which is distributed. Overall, in the case of no mode scrambling, the MDL STD increases linearly with the number of spans. In the case of strong mode scrambling, the MDL STD increases with the square root of the number of spans. 
These observations are consistent with the behavior of the MDL STD in the weak and strong random coupling regimes, respectively, in links with low overall MDL. 
In all cases with low overall MDL, the estimates from the analytical expression \eqref{eq:MDL_overall} closely match the numerical estimates from the simulation. 
The difference between the estimates for large $K$ in the case with no mode scrambling can be attributed to large overall MDL ($>13\,\text{dB}$), where the first-order approximation of the logarithm in \eqref{eq:gtot2_general} is not accurate.

{While we show the comparison of simulation and analytical results for only one fiber type, we have verified that the formulae are applicable to fibers with different numbers of modes or mode groups as long as they show strong intra-group coupling and weak inter-group coupling.}

\section{Design Criteria for a Strong Mode Scrambler}
\label{sec:Design_criteria}
In this section, we derive sufficient conditions on $\R$ to obtain GD and MDL STDs similar to those of systems with strong random mode coupling, wherein the STDs are proportional to the square root of the number of spans \cite{ho_linear_2014}.
Owing to our pessimistic assumption of no inter-group coupling in the MMFs, the estimates of GD and MDL STDs obtained here are typically higher than in practical scenarios with non-zero inter-group coupling in the MMFs. In view of the conservative assumption made, the obtained design criteria are robust.

As the closed-form expressions of GD STD \eqref{eq:gd_formula} and MDL STD \eqref{eq:MDL_overall} discussed in Section \ref{sec:GD_MDL_MS} depend on higher powers of $\DPC$, we are interested in the spectral properties of $\DPC$ and the behavior of $\left(\DPC\right)^k$ as $k\rightarrow\infty$. In this regard, the Perron-Frobenius theory for non-negative matrices \cite{meyer_matrix_2010} is quite useful.

\subsection*{Properties of $\mathbf{D}^{-1} \bm{\mathcal{P}}$}
Consider the eigenvalue decomposition $\DPC = \mathbf{U}\bm{\Lambda}\mathbf{U}^{-1}$, where $\bm{\Lambda} = \diag{\lambda_1,\dots,\lambda_{N_g}}$ is the diagonal matrix of eigenvalues and $\mathbf{U} = [\mathbf{u}_1,\dots,\mathbf{u}_{N_g}]$ is the matrix of right eigenvectors. 
Without loss of generality, $|\lambda_1| \geq |\lambda_2|\geq \dots \geq |\lambda_{N_g}|$. The spectral radius $\rho$ of $\DPC$ is defined as the maximum absolute value of the eigenvalues, $\rho(\DPC)=|\lambda_1|$.
The following are the properties of $\DPC$:
\begin{enumerate}
    \item $\DPC$ is a non-negative matrix, as is evident from the definition of $\PC$ in \eqref{eq:PC_def}. This means that the dominant eigenvalue is equal to the spectral radius; $\lambda_1 = \rho(\DPC)$.
    \item If the mode scrambler is power conserving, \textit{i.e.}, $\R$ is unitary, then:    
    \begin{enumerate}
        \item {$\DPC$ is a right stochastic matrix, defined as a non-negative square matrix, with each row summing to 1. $\PC \mathbf{D}^{-1}$ is a left stochastic matrix,defined as a non-negative square matrix, with each column summing to 1.}
        \item $\mathbf{1}_{N_g}$ is a right eigenvector of $\DPC$ with eigenvalue $\lambda_1=1$. $\mathbf{d}^T$ is the corresponding left eigenvector.
        \begin{equation*}
            \begin{split}
                \DPC \mathbf{1}_{N_g} ={}& \mathbf{1}_{N_g}\\
                \mathbf{d}^T \DPC ={}& \mathbf{d}^T.
            \end{split}
        \end{equation*}
        Similarly, $\mathbf{1}_{N_g}^T$ is a left eigenvector of $\PC \mathbf{D}^{-1}$ with eigenvalue $\lambda_1=1$. $\mathbf{d}$ is the corresponding right eigenvector:
        \begin{equation*}
            \begin{split}
                \mathbf{1}_{N_g}^T \PC \mathbf{D}^{-1}  ={}& \mathbf{1}_{N_g}^T\\
                 \PC \mathbf{D}^{-1} \mathbf{d}={}& \mathbf{d}.
            \end{split}
        \end{equation*}
        \item $\lambda_1=1$ is also the dominant eigenvalue of $\DPC$. The spectral radius, $\rho(\DPC)$ is therefore also equal to $1$.
    \end{enumerate}
\end{enumerate}

\subsection*{Design Criteria for $\DPC$}
{A practical mode scrambler device is not power-conserving and will exhibit MDL. However, a well-designed mode scrambler should have low intrinsic MDL. 
    Hence, the dominant eigenvalue of $\DPC$ will be close to $1$.}
The conditions on $\DPC$ for a mode scrambler such that the GD and MDL STDs are proportional to the square root of $K$ are:
\begin{enumerate}
    \item $\DPC$ is a primitive matrix. \newline
    {A matrix $\mathbf{B}\in \mathbb{C}^{n\times n}$ is said to be \textit{irreducible} if and only if for any two distinct indices $1\leq i,j \leq n$, there is a sequence of nonzero elements of $\mathbf{B}$ of the form $\{\mathbf{B}[i,i_1],\mathbf{B}[i_1,i_2],\dots,\mathbf{B}[i_m,j]\}$.
    $\mathbf{B}$ is said to be \textit{primitive} if it has real non-negative entries and there exists a positive integer $k$ for which every entry in $\mathbf{B}^k$ is positive.
    A primitive matrix is a special case of a non-negative irreducible matrix such that only one eigenvalue has an absolute value equal to its spectral radius.}
    It should be noted that:
    $$
    \PC\text{ is primitive} \Longleftrightarrow \DPC\text{ is primitive}.
    $$
    \item The non-dominant eigenvalues of $\DPC$ should be close to zero. 
\end{enumerate}


The conditions can be understood as follows.
In the first condition, the power coupling matrix implies all mode groups interact with each other in one or more passes through the mode scrambler. For example, in the simulations presented in Subsection~\ref{subsec:MultiSection_sims}, $\PC_0$ is a reducible matrix but $\PC_1$ and $\PC_2$ are irreducible matrices. As a result, the GD STD and the MDL STD corresponding to $\PC_0$ are higher than those corresponding to $\PC_1$ and $\PC_2$. 

Reducibility is a fundamental issue in the problem we are trying to solve. 
{
In the simulations in Subsection~\ref{subsec:MultiSection_sims}, weak inter-group coupling in the fiber spans leads to a reducible representation of the GDO and the MGO; each mode group can be considered an independent subsystem, and the mode-group-averaged delays and gains would drift apart as the number of spans increases.
As a result, the $D\times D$ GDO and MGO lead to expressions for the GD and MDL STDs involving $N_g \times N_g$ matrices. 
In the case of strong random coupling between all modes, the expressions could have been simplified to use $1\times 1$ scalar representations.}

$(\DPC)^{k}$ can be expanded as $\lambda_1^k \mathbf{u}_1 \acute{\mathbf{u}}_1^T + \dots + \lambda_{N_g}^k \mathbf{u}_{N_g} \acute{\mathbf{u}}_{N_g}^T$, where $\acute{\mathbf{u}}^T_1,\dots,\acute{\mathbf{u}}^T_{N_g}$ are the rows of $\mathbf{U}^{-1}$.
If $\DPC$ is primitive, then $(\DPC)^k$ tends to the rank-one matrix $\lambda_1^K\mathbf{u}_1 \acute{\mathbf{u}}_1^T$ as $k$ tends to infinity. Therefore, the contribution from the terms with $(\DPC)^k$ reduces as $k$ increases. 

The second condition is related to the convergence rate. If the non-dominant eigenvalues are close to zero, then $(\DPC)^k\approx\lambda_1^K\mathbf{u}_1 \acute{\mathbf{u}}_1^T$ is valid even for smaller values of $k$. The non-dominant eigenvalues of $\DPC_2$ are smaller in magnitude than those of $\DPC_1$. Hence, the GD and MDL STDs corresponding to $\PC_1$ are higher than those of $\PC_2$.

\subsection*{GD STD and MDL STD for a Strong Mode Scrambler}
We can show that GD and MDL STDs are proportional to $\sqrt{K}$ when $\DPC$ satisfies the above-mentioned conditions. In the low-MDL regime, $\lambda_1\approx1, \mathbf{u}_1\approx\mathbf{1}_{N_g}, \acute{\mathbf{u}}_1 \approx \mathbf{d}/D$. 
If $\DPC$ is primitive with small non-dominant eigenvalues,
\begin{equation*}
    (\DPC)^k\approx\lambda_1^k\mathbf{u}_1 \acute{\mathbf{u}}_1^T\approx  \frac{\lambda_1^k}{D} \mathbf{1}_{N_g} \mathbf{d}^T.
\end{equation*}
From \eqref{eq:gd_formula} we obtain
\begin{equation*}
    \begin{split}
        &\E{ \norm{\ttot}^2 }(K) \approx K\sum_{i=1}^{N_g} \sigma_{\text{intra},i}^2 + K \tautau_0^T \mathbf{D} \tautau_0 \\
    &+ \frac{1}{D}\tautau_0^T \left(
    2 \sum_{k=1}^{K-1} \left(K-k \right) \lambda_1^k \mathbf{1}_{N_g} \mathbf{d}^T \right)\tautau_0.
    \end{split}
\end{equation*}
Noting that $\mathbf{d}^T \tautau_0=\sum_{l=1}^{D}\beta_{1,l}=0$, we get
\begin{equation}
\label{eq:gd_formula_good_MS}
    \begin{split}
        &\E{ \norm{\ttot}^2 }(K) \approx K\sum_{i=1}^{N_g} \sigma_{\text{intra},i}^2 + K \tautau_0^T \mathbf{D} \tautau_0, \\
        \end{split}
\end{equation}
which is proportional to $K$. Therefore, $\sigma_{\mathrm{GD}}$ is proportional to $\sqrt{K}$.
Similarly, \eqref{eq:gtot2_approx} becomes
\begin{equation}
    \begin{split}
        \E{\norm{\gtotv}^2}(K)  \approx{}& 2\left( \frac{1}{D}\lambda_1^{K-1}  \mathbf{d}^T \mathbf{1}_{N_g} \mathbf{d}^T \mathbf{1}_{N_g}-D \right) \\
        ={}& 2D \left(\lambda_1^{K-1} -1 \right) \\
        \approx{}& 2D \left(K-1 \right) \left(\lambda_1 - 1\right), \\
    \end{split}
\end{equation}
which is proportional to $K$. Therefore, $\sigma_{\mathrm{MDL}}$ is proportional to $\sqrt{K}$.

\subsection*{Eigenvalue Space of $\DPC$}
\begin{figure}
\centering
\includegraphics[width=1\columnwidth]{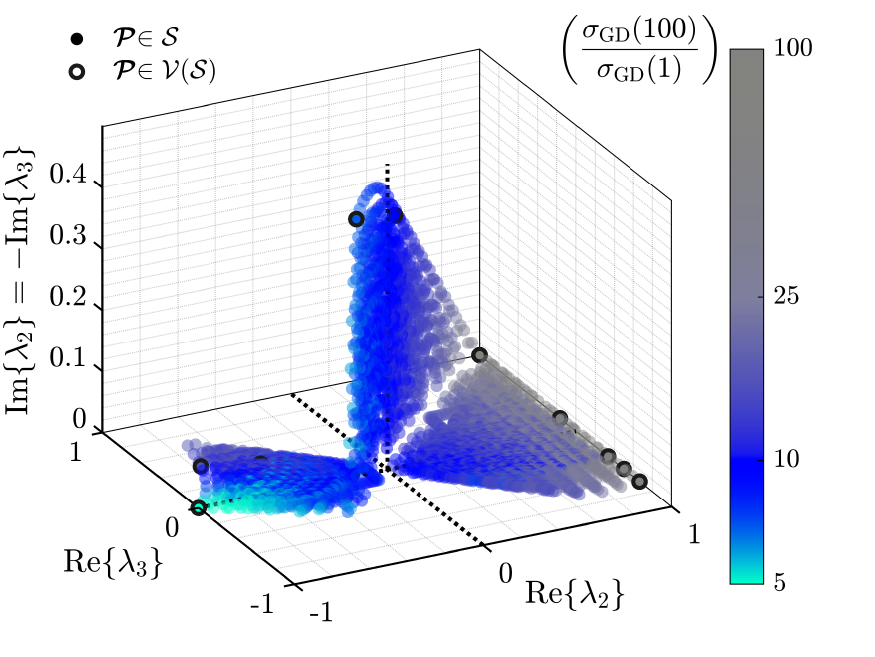}
\caption{Non-dominant eigenvalues of $\DPC$ for various $\PC$ and $\mathbf{D}=\diag{2,4,6}$. Each circle represents a different $\PC$ of size $3\times3$ from the set $\mathcal{S}$ such that $\DPC$ is right-stochastic and $\PC \mathbf{D}^{-1}$ is left-stochastic. The darkened circles represent $\PC$ matrices from the set of vertices $\mathcal{V}(\mathcal{S})$ of the convex hull $\mathcal{S}$. The color indicates the GD STD after $K=100$ spans of propagation normalized to $\sigma_{\mathrm{GD}}(1)$. The $\tautau_0$ parameter is taken from Table~\ref{tab:siml_params}.}
\label{fig:PC_eigs}
\end{figure}

Fig.~\ref{fig:PC_eigs} shows the dependence of the GD STD on the eigenvalues of $\DPC$ for $N_g = 3$, $\mathbf{d}=[2,4,6]^T$, and negligible MDL. The other simulation parameters have been taken from Table~\ref{tab:siml_params}.
We generate numerous $\PC$ matrices such that $\DPC$ is right stochastic and $\PC \mathbf{D}^{-1}$ is left stochastic, evaluate the formula in \eqref{eq:gd_formula} at $K=100$, and study the dependence of the normalized GD STD on the eigenvalues of $\DPC$. 
The set of all such matrices 
\begin{equation*}
\begin{split}
    \mathcal{S} =\biggl\{\PC \biggr\rvert &\PC\in\mathbb{R}^{N_g\times N_g}, \PC>0, \\
                                &\mathbf{D}^{-1} \PC \mathbf{1}_{N_g} = \mathbf{1}_{N_g}, \mathbf{1}_{N_g}^T \PC \mathbf{D}^{-1}  = \mathbf{1}_{N_g}^T \biggr\}
\end{split}
\end{equation*}
forms a convex hull. $\mathcal{V}(\mathcal{S})$ denotes the vertices of the convex hull.
The eigenvalues of the generated $\DPC$ are of the form $[\lambda_1 = 1, \lambda_2, \lambda_3]$ such that the non-dominant eigenvalues $\lambda_2$ and $\lambda_3$ are either complex conjugates of one another or both real.  
Fig.~\ref{fig:PC_eigs} is a scatter plot of $\lambda_2$ and $\lambda_3$. Each solid circle represents a $\PC$ from $\mathcal{S}$ and the color of the circle indicates the normalized GD STD $\sigma_{\mathrm{GD}}(K=100)/\sigma_{\mathrm{GD}}(1)$. 

The figure shows that the blue region represents strong scrambling with normalized GD STD close to $\sqrt{K}=10$. In this region, $\lambda_2$ and $\lambda_3$ are close to zero and the corresponding $\DPC$ matrices are dense and irreducible, making them suitable for scrambling. An example would be the all-ones matrix with the eigenvalues $\lambda_1 = 1$, $\lambda_2=0$, and $\lambda_3=0$. 
On the other hand, the gray region represents weak scrambling with normalized GD STD close to $K=100$. In this region, $\lambda_2$ and/or $\lambda_3$ are close to $1$ and the corresponding $\DPC$ matrices are not sufficiently dense and are not suitable for scrambling.
We also observe a cyan region around $(\lambda_2,\lambda_3)=(-1,0)$ that exhibits normalized GD STD values less than $\sqrt{K}=10$. This region corresponds to special $\PC$ matrices that satisfy $\tautau_0^T\PC \tautau_0<0$, or in other words, $\PC \tautau_0$ is anti-correlated with $\tautau_0$. This anti-correlation causes the arithmetico-geometric progression terms in \eqref{eq:gd_formula} to alternate in sign. These mode scramblers are ``self-compensating'', and are discussed in Section~\ref{subsec:GDcomp}.

\section{Effect of Inter-Group Coupling on GD STD}
\label{sec:Inter-Stokes}
In Section \ref{sec:GD_MDL_MS}, the inter-group coupling length was assumed to be infinite to obtain a closed-form expression of GD STD \eqref{eq:gd_formula} in the generalized Jones representation. 
While this assumption yields pessimistic estimates of GD STD and robust design criteria, it is important to quantify the impact of finite inter-group coupling that occurs in physical MMFs.
Hence, in this section, we remove this assumption and show that we can obtain a similar closed-form expression incorporating a finite inter-group coupling length for the case of $D=12$ modes and $N_g=3$ mode groups. 
The mathematical framework of the generalized Stokes space is preferred here as it conveniently incorporates random distributed coupling between any pair of modes, making it suitable for the analytical treatment of accumulated GD STD. 

Analysis in the generalized Stokes space involves representing the $D \times D$ GDO in the basis of $D^2-1$ generalized Pauli matrices \cite{antonelli_stokes-space_2012, arik_delay_2015}. 
This leads to stochastic differential equations governing the evolution of the $(D^2-1) \times 1$ modal dispersion vector in that basis. 
Subsequently, the expected modal dispersion vector and GD STD can be described by deterministic ordinary differential equations (ODEs).
Although these equations contain $(D^2-1)$-dimensional vectors and matrices, the GD STD can be well approximated by equations containing only $(D/2 - 1)$-dimensional vectors and matrices by assuming that polarization mode dispersion is negligible and that all mode coupling interactions are captured by appropriately chosen inter- and intra-group coupling lengths. 
Detailed analysis is presented in {\cite[Appendix D]{vijay_modal_2024}} and the closed-form expression for GD STD obtained from this analysis is similar in structure to the one derived in Section \ref{subsec:GDS}.
This enables us to obtain simple correction factors to include the effect of inter-group coupling in \eqref{eq:gd_formula}.
We verify all the resulting expressions using numerical modeling.

From the generalized Stokes analysis, we include simple correction factors in \eqref{eq:gd_formula} to capture finite inter-group coupling effects:
\begin{align}
    \begin{split}
        &\E{ \norm{\ttot}^2 }(K)  \\ &\approx K\sum_{i=1}^{N_g} \sigma_{\text{intra},i}^2   + K \frac{2\ztwo{\lenS}{\lenO/2}}{\lenS^2} \tautau_0^T \mathbf{D} \tautau_0  \\
    &+ \tautau_0^T \left( 2 \sum_{k=1}^{K-1} \left(K-k \right) \mathbf{D}\left(\DPC  \right)^{k} e^{\left(-\frac{2(k-1)\lenS}{\lenO} \right)} \right)\tautau_0 \\ &\times \left(\frac{\zone{\lenS}{\lenO/2}}{\lenS}\right)^2,
    \end{split}
    \label{eq:gd_formula_corrections}
\end{align}
where $\lenO$ is the inter-group coupling length and 
\begin{align*}
    \zone{z}{L} &= L\left(1-\exp{\left(-\frac{z}{L}\right)} \right),\\
    \ztwo{z}{L} &= Lz-L^2\left(1-\exp{\left(-\frac{z}{L}\right)} \right).
\end{align*}

\begin{figure}
\centering
\includegraphics[width=1\columnwidth]{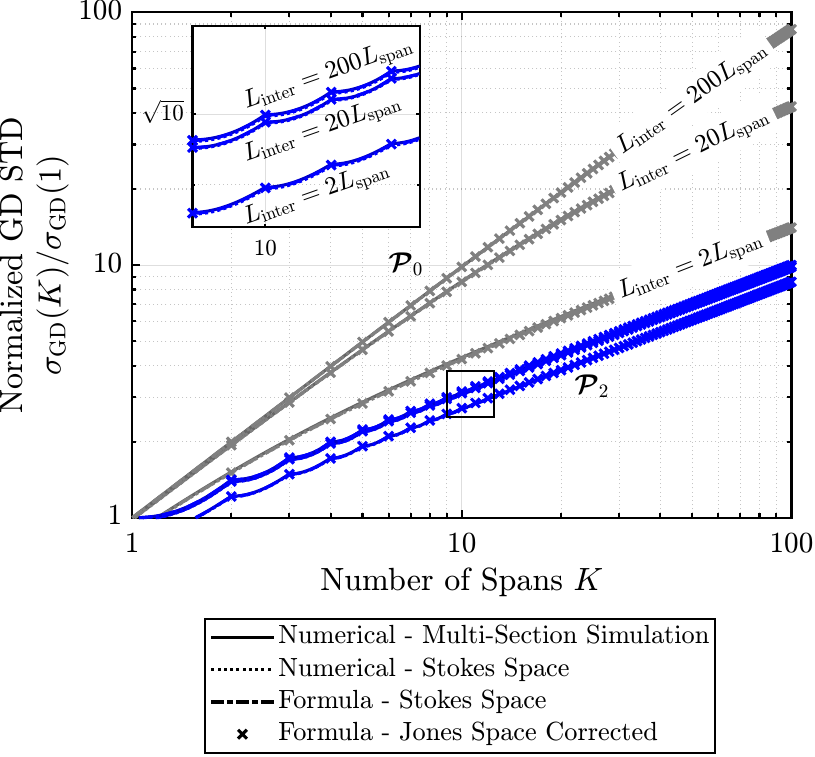}
\caption{GD STD as a function of the number of spans for various inter-group coupling lengths and two scrambling scenarios. The solid lines correspond to numerical estimates from multi-section simulations, the dotted lines correspond to the numerical evaluation of the two ODEs {\cite[eq. S29]{vijay_modal_2024}} and {\cite[eq. S30]{vijay_modal_2024}} from the Stokes analysis, the dashed lines correspond to the analytical formula from the Stokes analysis {\cite[eq. S41]{vijay_modal_2024}}, and the crosses correspond to the analytical formula from the Jones analysis with corrections for a finite inter-group coupling length (\ref{eq:gd_formula_corrections}). The GD STD $\sigma_{\mathrm{GD}}(K)$ is normalized by the one-span GD STD $\sigma_{\mathrm{GD}}(1)$.}
\label{fig:stokes_GDS_sims}
\end{figure}

Fig. \ref{fig:stokes_GDS_sims} shows the normalized GD STD $\sigma_{\mathrm{GD}}(K)/\sigma_{\mathrm{GD}}(1)$ plotted as a function of the number of spans $K$ for $\lenO \in \{2\lenS$, $ 20\lenS$, $ 200\lenS\}$. 
We consider power coupling matrices $\PC_0$ and $\PC_2$ corresponding to the cases with no scrambling and strong scrambling, respectively, as in Section \ref{subsec:MultiSection_sims}. 
To isolate the effects of inter-group coupling, we consider no MDL in the mode scrambler transfer matrices. 
The plot compares the GD STDs obtained by four methods: the multi-section model, numerical evaluation of the two ODEs {\cite[eq. S29]{vijay_modal_2024}} and {\cite[eq. S30]{vijay_modal_2024}}, and the analytical estimates from the Stokes formula {\cite[eq. S41]{vijay_modal_2024}} and from the Jones formula with corrections for finite inter-group coupling length \eqref{eq:gd_formula_corrections}. 
We see good agreement between the analytical formulae and the estimates from numerical simulation.

\section{Equivalence to Strong Random Coupling}
\label{sec:strong_random_coupling}

In deriving the conditions for a strong mode scrambler in Section \ref{sec:Design_criteria}, we assumed that the mode scramblers with deterministic transfer matrices are the only source of inter-group coupling and that the fiber sections are a source of random intra-group coupling. 
We showed that such a design can replicate the effect of strong random coupling between all modes by analyzing GD and MDL STDs. 
The question of whether GD and MDL STDs expressions in Section~\ref{sec:GD_MDL_MS} are definitive indicators of system performance is yet unanswered {in this paper}.
However, as a result of the Central Limit Theorem (CLT), by combining strong deterministic mode scrambling with strong intra-group coupling, we expect the probability distributions of modal delays and modal gains to approach those in systems with strong random coupling between all modes.

In this section, we (i) demonstrate close agreement between the tail probabilities of the peak-to-peak GD spread and the cumulative distribution function (CDF) of the capacity of two systems: one with strong random coupling between all modes and the other with mode-scrambler-aided coupling satisfying the design criteria in Section \ref{sec:Design_criteria}, and
(ii) discuss the asymptotic convergence of the distributions of the system transfer matrices in both cases.

\subsection{Group-Delay Tail Probabilities and Distribution of\\Capacity}
In terms of system complexity and performance, two distributions are particularly useful, namely, the tail probabilities of the peak-to-peak GD spread and the cumulative distribution function (CDF) of the capacity \cite{kaminow_chapter_2013}. 
The DSP complexity of the receiver equalizers depends on the statistics of the peak-to-peak GD spread.
The average and outage capacities can be computed from the capacity distribution, which depends on the distribution of modal power gains.
In the presence of strong random coupling between all modes, the distributions of the group delays and the modal gains are well known, and the GD STD and the MDL STD fully describe the respective distributions \cite{ho_linear_2014}. Moreover, the peak-to-peak GD spread and the GD STD are equivalent metrics. 
Using multi-section simulations, we verify here that these distributions are similar for both strong random coupling and strong deterministic mode scrambling with strong intra-group coupling, thereby validating the applicability of the formulae in Section~\ref{sec:GD_MDL_MS}.

\begin{figure}
\centering
\includegraphics[width=1\columnwidth]{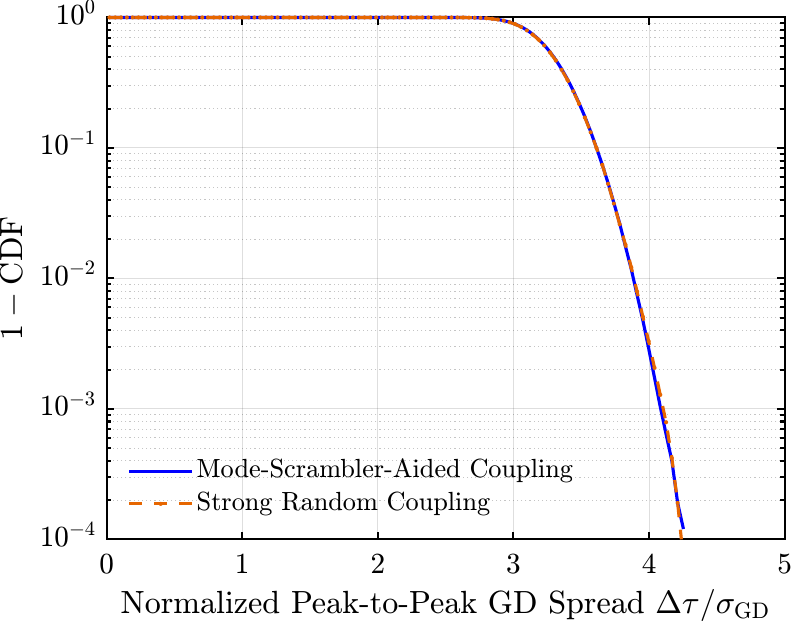}
\caption{Complementary CDF of the peak-to-peak GD spread normalized to the GD STD after $K=100$ spans of propagation. The blue solid line corresponds to mode-scrambler-aided deterministic coupling for $D=12, N_g = 3, \mathbf{D}=\diag{2,4,6}$ and the orange dashed line corresponds to strong random coupling for $D=12$.}
\label{fig:tail_GDS}
\end{figure}

Fig.~\ref{fig:tail_GDS} shows the complementary CDF of the normalized peak-to-peak GD spread after $K=100$ spans of propagation for mode-scrambler-aided coupling and strong random coupling. The figure highlights the tail probabilities of the peak-to-peak GD spread. The mode scrambler transfer matrix is chosen according to the design criteria in \ref{sec:Design_criteria}. To obtain the CDF, random fiber realizations are generated and the peak-to-peak GD spread is computed as the difference between the maximum and the minimum GDs, $\Delta\tau = \max_l(\tau_{\textrm{tot},l}) - \min_l(\tau_{\textrm{tot},l})$. The quantity is then normalized by $\sigma_{\mathrm{GD}}$. 

\begin{figure}
\centering
\includegraphics[width=1\columnwidth]{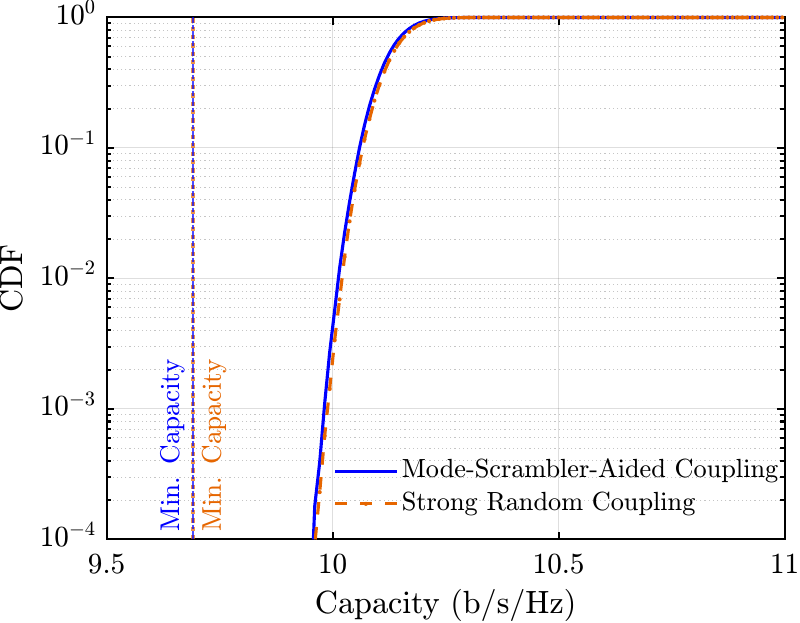}
\caption{CDF of capacity after $K=100$ spans of propagation. The blue solid line corresponds to mode-scrambler-aided deterministic coupling for $D=12, N_g = 3, \mathbf{D}=\diag{2,4,6}$ and the orange dashed line corresponds to strong random coupling for $D=12$, for an SNR of $10~\text{dB}$. The vertical dotted lines indicate the minimum capacities.}
\label{fig:capacity}
\end{figure}

Fig.~\ref{fig:capacity} shows the CDF of the capacity after $K=100$ spans of propagation for mode-scrambler-aided coupling and strong random coupling for a signal-to-noise ratio (SNR) of $10~\text{dB}$. 
The per-span MDL STD $\sigma_{\mathrm{MDL},0}$ is set to $0.2$~dB. 
The channel capacity, measured in $\mathrm{b/s/Hz}$, assuming channel-state information is not available, is defined as \cite{mello_impact_2020}
\begin{equation*}
    C = \log_2 \det \left( \mathbf{I}_D - \frac{\mathrm{SNR}}{D}\widehat{\mathbf{M}}_\textrm{tot} \widehat{\mathbf{M}}^H_\textrm{tot}\right),
\end{equation*}
where $\widehat{\mathbf{M}}_\textrm{tot} = \mathbf{M}_\textrm{tot}/\alpha_0$, and the SNR is in linear units. $\alpha_0$ is the ensemble-averaged mode-averaged gain, and $\alpha_0 = \sqrt{\sum_{l=1}^D \E{e^{\gtot{l}}}/D}$. 
Capacity is lower bounded by $C_\textrm{min} = D \log_2 \left(1 + \mathrm{SNR}/\alpha_0^2 D \right)$ \cite{mello_impact_2020}.

Fig.~\ref{fig:tail_GDS} and Fig.~\ref{fig:capacity} are consistent with the assumption that combining periodic deterministic scrambling with random intra-group coupling results in the same distribution of modal delays and modal gains as a system with strong random coupling between all modes.

\subsection{Asymptotic Convergence of the Distribution of the System Transfer Matrix}

Transfer matrices of fiber sections with strong random coupling can be modeled by random unitary matrices distributed according to the Haar measure \cite{kaminow_chapter_2013}. 
In a system with strong random coupling between all modes, wherein each fiber section effects an independent random unitary transformation, the GDO can be written as a sum of section-wise GDOs with IID eigenvectors. As a result of the CLT, the overall GDO asymptotically approaches a Gaussian Unitary Ensemble (GUE) with zero trace.

In a system with deterministic mode scrambling and strong random intra-group coupling, we expect that the system transfer matrix asymptotically approaches, in distribution, a random unitary matrix distributed according to the Haar measure.
As a result, the GDO can be written as a sum of the span-wise GDOs with asymptotically IID eigenvectors.
Here too, the CLT holds and the overall GDO asymptotically approaches a GUE with zero trace.

Arguments supporting the convergence of the distribution of $\mathbf{M}_\textrm{tot}$ according to the Haar measure on the unitary group in the presence of strong mode scrambling, and the convergence of $\mathbf{G}_\textrm{tot}$ to a GUE is provided in {\cite[Appendix G]{vijay_modal_2024}}.
We pessimistically assume no inter-group coupling in the fiber.
For simplicity, we also assume that the mode scrambler transfer matrix $\R$ is unitary.
We define a Markov Chain on the system transfer matrix with the number of spans $K$ as the Markov ``time-step''.
In this framework, the system transfer matrix is a continuous-valued Markov ``state" on the compact topological group of $D\times D$ unitary matrices.
We argue that this Markov chain has key properties like irreducibility, aperiodicity, and recurrence when $\R$ satisfies the design criteria in Section \ref{sec:Design_criteria}, and consequently establishes that the Law of Large Numbers (LLN) and the CLT hold for a series summation of certain classes of functions of the Markov state, including the GDO.

The eigenvalue distribution of a GUE with zero trace is well known \cite{kaminow_chapter_2013}.
The asymptotic convergence of the GDO therefore establishes that the mean and variance of the eigenvalues are sufficient statistics
The mean is zero owing to the zero-trace property and the variance is given by \eqref{eq:gd_formula}.
While we assumed that MDL is not present, it can be argued that when the overall MDL is low, the distribution of the eigenvalues of the MGO (measured in decibels) converges to a GUE with zero trace.
Although the analysis in {\cite[Appendix G]{vijay_modal_2024}} indicates asymptotic convergence, our simulations demonstrate a rapid rate of convergence to the limiting distributions.
This is supported by the close agreement shown in Fig. \ref{fig:tail_GDS} and Fig. \ref{fig:capacity} for $K = 100$ spans.

\section{Discussion}
\label{sec:discussion}
In this section, we discuss some aspects of long-haul system design with mode scramblers.


\subsection{Non-Identical Mode Scramblers}

The derivation of the analytical expressions in Section~\ref{sec:GD_MDL_MS} assumed that all mode scramblers have identical transfer matrices.
While this scenario is pessimistic in terms of mode-scrambling performance, it leads to robust design criteria.
In a practical system, the mode scramblers could have different transfer matrices by design or due to manufacturing variations.
In such cases, the formulae can be modified to incorporate the different power coupling matrices.
In both cases, if individual mode scrambler devices satisfy the design criteria, then the GD and MDL STDs will resemble those of a system with strong mode coupling.
A system design can be verified by analytical modeling using the measured power coupling matrices of fabricated mode scramblers.





\subsection{Spatial Whiteness of Amplifier Noise}
Spatial whiteness of the total amplified spontaneous emission (ASE) noise is another consequence of strong mode scrambling.
The ASE noises from the amplifiers are not individually spatially white owing to MDL. 
The $D-$dimensional noise vectors from each amplifier experience mode scrambling and transformation by the fiber matrices before they reach the receiver. 
Following the analysis in \cite[Discussion]{ho_mode-dependent_2011}, the noise correlation matrices can be written in terms of the mode-dependent noise variances of each amplifier, the fiber transfer matrices $\{\mathbf{M}_k\}$ and $\R$, which has a form very similar to the terms in the expansion of the GDO {\cite[eq. S3]{vijay_modal_2024}}.
The equations are provided in {\cite[Appendix H]{vijay_modal_2024}}.
Equivalently, we can write the noise correlation matrices as a function of the Markov state.
Their expected values are diagonal matrices and can be computed similarly to the analysis in {\cite[Appendix B]{vijay_modal_2024}}.
When the number of spans is large, owing to the LLN, we can argue that the overall noise correlation matrix converges to a constant times the identity matrix; in other words, the noise is spatially white.

\subsection{Group-Delay Self-Compensation}
\label{subsec:GDcomp}

A good design for the mode scrambler does not necessarily solve all system-level problems.
Strong mode scrambling is equivalent to strong random coupling with an effective inter-group coupling length of $\lenS$: $\sigma_{\mathrm{GD}}(K) \approx \sigma_{\beta_1}\sqrt{K\lenS L_{\textrm{inter,eff}}}$, where $\sigma_{\beta_1} = \sqrt{(\beta_{1,1}^2+\dots+\beta_{1,D}^2)/D}$ and $L_{\textrm{inter,eff}} = \lenS$.
For optimum performance, the transmission fiber must have low uncoupled GD STD $\sigma_{\beta_1}$, and the optical devices must have low intrinsic MDL.
Since modal dispersion is a distributed effect, placing multiple mode scramblers between successive amplifiers can be beneficial as it reduces the multiplicative factor $\sqrt{L_{\textrm{inter,eff}}}$ in $\sigma_{\mathrm{GD}}$. However, there is a trade-off: the overall MDL will increase as more mode scramblers are used.
Another approach is to co-design mode scramblers with transmission fibers. This can potentially result in $L_{\textrm{inter,eff}}$ being less than $\lenS$. This corresponds to the cyan region in Fig.~\ref{fig:PC_eigs}, where the power coupling matrix is such that $\PC\tautau_0$ is anti-correlated with $\tautau_0$ and exhibits GD compensating behavior. 

Scrambling to reduce the GD spread involves frequently splitting and redirecting data signals to different modes such that no data signal consistently travels in the fastest or the slowest mode.
An optimal solution can be obtained by carefully redirecting data signals such that they transfer between propagation in fast and slow modes. This results in a net delay spread that is periodically minimal. 
This forms the basis of GD compensation using multiple transmission fiber types \cite{randel_mode-multiplexed_2012,arik_delay_2015,bai_equalizer_2014}. 
GD compensation can also be performed by careful mode permutations. 
In Fig.~\ref{fig:PC_eigs}, the power coupling matrix corresponding to the minimum GD STD is
\begin{equation*}
    \PC_3 = \begin{bmatrix}
        0 & 0 & 2 \\
        0 & 0 & 4 \\
        2 & 4 & 0
    \end{bmatrix},
\end{equation*}
which routes all the power from the first and second mode groups to the third mode group and vice-versa.
We can see intuitively that this is the best option because the average group delays of the first and second mode groups are positive, while the average group delay of the third mode group is negative, $\tautau_0 = [25.2, 7.05,$ $-13.1]^T\,\text{ns}$.
A matrix $\R_3$ with the above power coupling matrix is the anti-diagonal Exchange Matrix $\mathbf{J}_{12}$ whose elements are $1$ on the anti-diagonal and $0$ otherwise. 
The appropriate term for these devices is ``mode permutators'', as they do not scramble the modes. 
The technique of mode permutation has been proposed in the literature for the management of GD spread \cite{shibahara_long-haul_2020,di_sciullo_reduction_2023, hout_transmission_2024} and MDL caused by splices \cite{warm_splice_2013}. 

\begin{figure}
\centering
\includegraphics[width=1\columnwidth]{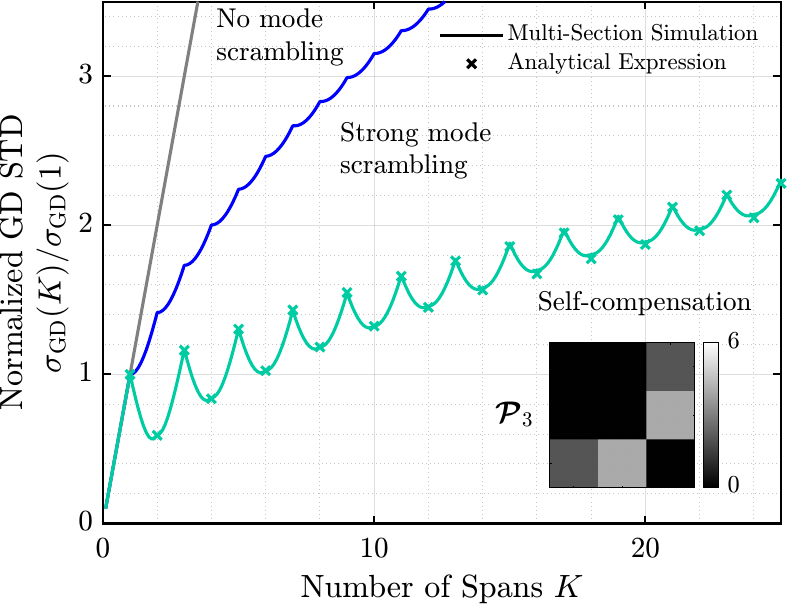}
\caption{Self-compensation with a mode permutator for $\mathbf{D}=\diag{2,4,6}$. The solid lines correspond to numerical estimates from multi-section simulations and the x-markers correspond to analytical estimates from the formula in \eqref{eq:gd_rms_tot}. GD STD $\sigma_{\mathrm{GD}}(K)$ is normalized by one-span GD STD $\sigma_{\mathrm{GD}}(1)$. The inset shows the power coupling matrix $\PC_3$ used in obtaining the self-compensation curve.}
\label{fig:comp_gds}
\end{figure}

Fig.~\ref{fig:comp_gds} shows the normalized GD STD plotted as a function of the number of spans $K$ for a self-compensating mode permutator with the above power coupling matrix. 
Similarly to strong scrambling, the GD STD for the self-compensating mode permutator increases with the square root of the number of spans but $L_{\textrm{inter,eff}}$ is less than $\lenS$.
The alternating pattern in the GD STD is expected as a result of the anti-correlation of $\PC_3\tautau_0$ and $\tautau_0$.
$\DPC_3$ has the eigenvalues $[\lambda_1 = 1, \lambda_2 = -1, \lambda_3 = 0]$, and satisfies the following property: $(\DPC_3)^k=\DPC_3$ when $k$ is odd and $(\DPC_3)^k=(\DPC_3)^2$ when $k$ is even. Therefore, \eqref{eq:gd_formula} for even $K$ becomes,
\begin{equation*}
\label{eq:gd_formula_comp_MS}
    \begin{split}
        &\E{ \norm{\ttot}^2 }(K) \approx \\
        &\left( K\sum_{i=1}^{N_g} \sigma_{\text{intra},i}^2 + K\tautau_0^T \mathbf{D}\left(\mathbf{I}_{N_g} - (\DPC_3)^2 \right)\tautau_0\right),
    \end{split}    
\end{equation*}
which is also proportional to $K$, similar to the expression for GD STD for a strong mode scrambler in \eqref{eq:gd_formula_good_MS}. 
However, there is an extra factor of $\mathbf{I}_{N_g} - (\DPC)^2$ compared to \eqref{eq:gd_formula_good_MS}; therefore $L_{\textrm{inter,eff}}<\lenS$. 

It should be noted that $\DPC_3$ is an irreducible matrix but not primitive. Hence, it does not satisfy the conditions of a strong mode scrambler for arbitrary $\tautau_0$. For a different transmission fiber, the optimum power coupling matrix may differ from $\PC_3$, especially if the signs of $\tautau_0$ change. We can define an optimization problem to obtain the best power coupling matrix for any given $\tautau_0$ by minimizing \eqref{eq:gd_formula} for even and large $K$. We start by minimizing \eqref{eq:gd_formula} for $K=2$ spans by formulating the following optimization problem 
\begin{equation}
    \label{eq:comp_gds_optim}
    \begin{aligned}
        \minimize_{\PC}     &\quad \tautau_0^T \PC \tautau_0\\
        \subjectto          &\quad \PC[i,j] \geq 0,~1\leq i,j \leq N_g \\
                            &\quad \DPC \mathbf{1}_{N_g} = \mathbf{1}_{N_g}\\
                            &\quad \mathbf{d}^T \DPC = \mathbf{d}^T\\
    \end{aligned},
\end{equation}
where the constraints enforce that $\PC$ is a non-negative matrix and $\R$ is a unitary matrix. The constraints are convex and the cost function is linear in $\PC$. 
Therefore, we have a linear programming problem. The constraint set is feasible. It is well-known that the optimum value in a linear programming problem is found on the boundary of the constraint set \cite{chvatal_linear_1983}. To obtain an optimum $\DPC$, we need to only search over the set of vertices of the convex hull $\mathcal{S}$. If multiple vertices attain the same minimum value of the cost function, then any convex combination of those vertices is a valid optimum.

An optimum $\PC$ obtained for the minimization problem in \eqref{eq:comp_gds_optim} does not necessarily minimize the GD STD for $K>2$ spans, wherein the cost function is non-convex in $\PC$. However, we have verified through gradient descent searches in the constraint set that the optimum $\PC$ for $K=2$ and $K>2$ are identical for $\tautau_0$ values similar to the one in Table~\ref{tab:siml_params}.

An important challenge for this approach lies in realizing a device with an optimal power coupling matrix. Possible device solutions include LPFGs and back-to-back multiplexer-demultiplexer using photonic lanterns, which require further study.
To obtain maximum benefit, it is also necessary to carefully design the transmission fiber to yield a favorable $\tautau_0$.
Using mode permutators for self-compensation has limitations caused by random inter-group mode coupling, similar to GD compensation using different fibers \cite{arik_group_2016}.
The GD STD derived from the Stokes space analysis {\cite[eq. S41]{vijay_modal_2024}} can be useful in evaluating the performance of systems designed with this technique.
While this technique optimizes GD STD, its effect on MDL STD needs further study.
These are topics for future research.

\section{Conclusion}
\label{sec:conclusion}
In this paper, we studied the effects of the mode scrambler transfer matrix on modal dispersion and mode-dependent gain/loss. 
Using a generalized Jones representation, we derived analytical expressions for GD STD and overall MDL STD in terms of the mode-group power coupling matrix for systems with strong random intra-group coupling and weak inter-group coupling in the MMF. 
Through multi-section simulations, we verified the analytical expressions in the low-MDL regime.
We also proposed the following design criteria for a mode scrambler to obtain GD and MDL STDs proportional to the square root of the number of spans of propagation: the power coupling matrix $\PC$ should be primitive and the non-dominant eigenvalues of $\DPC$ should be close to zero. Systems using such mode scramblers exhibit inter-group coupling lengths equal to the span length $\lenS$.
Using a generalized Stokes representation, we re-derived analytical expressions for GD STD to incorporate the effects of inter-group coupling in the MMF.

We argued that deterministic mode scrambling in the presence of strong intra-group coupling is asymptotically equivalent to strong random coupling.
We also verified that the tail probabilities of peak-to-peak GD spread and channel capacity match in the two scenarios.
Finally, we discussed the self-compensation regime wherein the GD STD periodically increases and decreases and has an effective inter-group coupling length less than $\lenS$.

\appendices

\section*{Acknowledgment}
This project was supported by Ciena Corporation and a Stanford Shoucheng Zhang Graduate Fellowship.
Much of the computing for this project was performed on the Sherlock cluster at Stanford University. We thank the Stanford Research Computing Center for providing this cluster and technical support.
We are grateful for helpful discussions with Prof.~David Miller, Prof.~Shanhui Fan, and Prof.~Joonhee Choi.

\bibliographystyle{IEEEtran}
\bibliography{references}

\begin{IEEEbiographynophoto}{Anirudh Vijay}
received the B.Tech. and M.Tech. degrees in Electrical Engineering from the Indian Institute of Technology Madras, Chennai, Tamil Nadu, India, in 2019.
He is working towards the Ph.D. degree in Electrical Engineering from Stanford University, Stanford, CA, USA. 
His current research interests include optical communications, mode-division multiplexing, and data-center applications.
\end{IEEEbiographynophoto}

\begin{IEEEbiographynophoto}{Oleksiy Krutko}
received the B.S. degree in electrical engineering from the University of Texas at Austin, Austin, TX, USA, in 2020. He is currently working toward the Ph.D. degree from Stanford University, Stanford, CA, USA. His research interests include optical fiber communications and photonic devices.
\end{IEEEbiographynophoto}

\begin{IEEEbiographynophoto}{Rebecca Refaee}
received the B.S. degree in mathematics and the M.S. degree in electrical engineering from Stanford University, Stanford, CA, USA in 2024. 
She is currently working towards the Ph.D. degree in electrical engineering at Stanford University. Her current research interests include optical communications and mode-division multiplexing.
\end{IEEEbiographynophoto}

\begin{IEEEbiographynophoto}{Joseph M. Kahn}
(F’00) received A.B., M.A. and Ph.D. degrees in Physics from the University of California, Berkeley in 1981, 1983 and 1986. In 1987-1990, Kahn was at AT\&T Bell Laboratories. In 1989, he demonstrated the first successful synchronous (i.e., coherent) detection using semiconductor lasers, achieving record receiver sensitivity. In 1990-2003, Kahn was on the Electrical Engineering and Computer Sciences faculty at Berkeley. He demonstrated coherent detection of QPSK in 1992. In 1999, D. S. Shiu and Kahn published the first work on probabilistic shaping for optical communications. In the 1990s and early 2000s, Kahn and collaborators performed seminal work on indoor and outdoor free-space optical communications and multi-input multi-output wireless communications. In 2000, Kahn and K. P. Ho founded StrataLight Communications, whose 40 Gb/s-per-wavelength long-haul fiber transmission systems were deployed widely by AT\&T, Deutsche Telekom, and other carriers. In 2002, Ho and Kahn applied to patent the first electronic compensation of fiber Kerr nonlinearity. StrataLight was acquired by Opnext in 2009. In 2003, Kahn became a Professor of Electrical Engineering in the E. L. Ginzton Laboratory at Stanford University. Kahn and collaborators have extensively studied rate-adaptive coding and modulation, as well as digital signal processing for mitigating linear and nonlinear impairments in coherent systems. In 2008, E. Ip and Kahn (and G. Li independently) invented simplified digital backpropagation for compensating fiber Kerr nonlinearity and dispersion. Since 2004, Kahn and collaborators have studied propagation, modal statistics, spatial multiplexing and imaging in multimode fibers, elucidating principal modes and demonstrating transmission beyond the traditional bandwidth-distance limit in 2005, deriving the statistics of coupled modal group delays and gains in 2011, and deriving resolution limits for imaging in 2013. Kahn’s current research addresses optical frequency comb generators, coherent data center links, rate-adaptive access networks, fiber Kerr nonlinearity mitigation, ultra-long-haul submarine links, and optimal free-space transmission through atmospheric turbulence. Kahn received the National Science Foundation Presidential Young Investigator Award in 1991. In 2000, he became a Fellow of the IEEE.
\end{IEEEbiographynophoto}

\title{Supplementary Material: Modal Statistics in Mode-Division-Multiplexed Systems\\using Mode Scramblers}
\maketitle

\appendices

\section{Expectation Over the Group of Block Unitary Matrices}
In this appendix, we analytically compute $\E{\mathbf{T}_\textrm{block} \mathbf{A} \mathbf{T}^H_\textrm{block}}$, where $\mathbf{A}$ is a deterministic matrix and $\mathbf{T}_\textrm{block} = \diag{\mathbf{T}_{11}, \mathbf{T}_{22}, \dots, \mathbf{T}_{N_gN_g}}$ is a block-diagonal random unitary matrix whose $i$th block diagonal matrix $\mathbf{T}_{ii}$ is independently distributed according to the Haar measure on the group of $d_i\times d_i$ unitary matrices.
The $i,j$ submatrix of $\E{\mathbf{T}_\textrm{block} \mathbf{A} \mathbf{T}^H_\textrm{block}}$ can be written as
\begin{align}
    \begin{split}
            \E{\mathbf{T}_\textrm{block} \mathbf{A} \mathbf{T}^H_\textrm{block}}_{ij} ={}& \E{\mathbf{T}_{ii} \mathbf{A}_{ij} \mathbf{T}^H_{jj}} \\
            ={}&
            \begin{cases}
            \mathbf{0}_{d_i\times d_j}, & i\neq j, \\
            \frac{1}{d_i}\tr{\mathbf{A}_{ii}} \mathbf{I}_{d_i}, & i=j,
            \end{cases}       
    \end{split} \label{eq:Ublock}
\end{align}
where $\mathbf{0}_{d_i\times d_j}$ is the $d_i\times d_j$ zero matrix.
The result for $i=j$ in \eqref{eq:Ublock} comes from the integration with respect to the Haar measure on the unitary group \cite{collins_integration_2006}.
Alternatively, in terms of $\{\bm{\Gamma_i} \}$ matrices, we can write $\E{\mathbf{T}_\textrm{block} \mathbf{A} \mathbf{T}^H_\textrm{block}}$ as
\begin{align}
    \label{eq:Uni_exptn_block}
    \E{\mathbf{T}_\textrm{block} \mathbf{A} \mathbf{T}^H_\textrm{block}} ={}& \sum_{i=0}^{N_g}\frac{1}{d_i}\tr{\mathbf{A}_{ii}}\bm{\Gamma}_{i}\nonumber \\
    ={}&\sum_{i=0}^{N_g}\frac{1}{d_i}\tr{\bm{\Gamma}_{i}\mathbf{A}}\bm{\Gamma}_{i}.
\end{align}
\section{Derivation of GD STD in the Generalized Jones Representation}
\label{app:RMS_GDS}
In this appendix, we provide the derivation of $\E{\norm{\ttot}^2 }$.
From the definitions of $\mathbf{M}_{\textrm{tot}}$ and $\mathbf{G}_{\textrm{tot}}$, and assuming $\mathbf{M}_{\textrm{tot}}$ is unitary, we get

\begin{align}
    \begin{split}
        &\mathbf{G}_{\textrm{tot}}   ={} -\jmath \mathbf{M}_{\textrm{tot}}^H \frac{\partial\mathbf{M}_{\textrm{tot}}}{\partial\omega} = \jmath \frac{\partial\mathbf{M}^H_{\textrm{tot}}}{\partial\omega}\mathbf{M}_{\textrm{tot}} 
    \end{split}\nonumber\\
    \begin{split}\label{eq:G_tot_expansion}
        &={} \mathbf{G}_{1} + \mathbf{M}^H_{1} \R^{H} \mathbf{G}_{2} \R \mathbf{M}_{1} + \dots + \\
         &\quad{} \mathbf{M}^H_{1} \R^{H} \dots \mathbf{M}^H_{K-1} \R^{H} \mathbf{G}_{K} \R \mathbf{M}_{K-1} \dots \R \mathbf{M}_{1},
    \end{split}                      
\end{align}
where $\mathbf{G}_k = -\jmath \mathbf{M}^{H}_k \frac{\partial\mathbf{M}_k}{\partial\omega}$ is the GDO of the $k$th span for $k=1,\dots,K$. Using the expansion of $\mathbf{G}_{\textrm{tot}}$ in \eqref{eq:G_tot_expansion}, we obtain $\E{\norm{\ttot}^2 }$ as a sum of $K^2$ terms, as shown in \eqref{eq:tau_tot}. Assuming that each fiber span is statistically identical, \eqref{eq:tau_tot} simplifies to \eqref{eq:gd_formula_baby}.
\begin{figure*}[ht!]
   \hrulefill
    \begin{align}
    \E{\norm{\ttot}^2 } &= \E{ \tr{\mathbf{G}^H_{\textrm{tot}} \mathbf{G}_{\textrm{tot}}} } \nonumber \\
    \begin{split}
        &= \biggl({}\sum_{k=1}^{K} \E{ \tr{\mathbf{G}^H_{k} \mathbf{G}_{k}} } \\
        &\qquad+  2 \sum_{k=1}^{K-1} \E{ \tr{\mathbf{G}^H_{k+1} \R \mathbf{M}_{k} \mathbf{G}_{k} \mathbf{M}^H_{k} \R^{H} } } \\
        &\qquad+  2 \sum_{k=1}^{K-2} \E{ \tr{\mathbf{G}^H_{k+2} \R \mathbf{M}_{k+1} \R \mathbf{M}_{k} \mathbf{G}_{k} \mathbf{M}^H_{k} \R^{H} \mathbf{M}^H_{k+1} \R^{H} } } \\
        &\qquad+ \dots \\
        &\qquad+  2{}\E{ \tr{\mathbf{G}^H_{K} \R \mathbf{M}_{K-1} \dots \R \mathbf{M}_{1} \mathbf{G}_{1} \mathbf{M}^H_{1} \R^{H} \dots \mathbf{M}^H_{K-1} \R^{H} } }\biggr).
    \end{split}\label{eq:tau_tot}\\
    {}\E{ \norm{\ttot}^2 }(K) &= \biggl( K \E{ \tr{\mathbf{G}^H_{1} \mathbf{G}_{1}} } \nonumber \\
    \begin{split}
        &\qquad+  2 (K-1) \E{ \tr{\mathbf{G}^H_{2} \R \mathbf{M}_{1} \mathbf{G}_{1} \mathbf{M}^H_{1} \R^{H} } } \\
        &\qquad+  2(K-2) \E{ \tr{\mathbf{G}^H_{3} \R \mathbf{M}_{2} \R \mathbf{M}_{1} \mathbf{G}_{1} \mathbf{M}^H_{1} \R^{H} \mathbf{M}^H_{2} \R^{H} } } \\
        &\qquad+  \dots \\
        &\qquad+  2 \E{ \tr{\mathbf{G}^H_{K} \R \mathbf{M}_{K-1} \dots \R \mathbf{M}_{1} \mathbf{G}_{1} \mathbf{M}^H_{1} \R^{H} \dots \mathbf{M}^H_{K-1} \R^{H} } }\biggr). \label{eq:gd_formula_baby}
    \end{split}
    \end{align} 
    \hrulefill
\end{figure*}

To obtain $\E{\norm{\ttot}^2 }$, we find analytical expressions for each term in \eqref{eq:gd_formula_baby}.

\subsection*{Analytical Expression for $\E{ \mathbf{G}_{k} }$}
Assuming strong intra-group coupling and negligible inter-group coupling, for $k=1,2,\dots,K$, $\mathbf{M}_{k}$ and $\mathbf{G}_{k}$ are block diagonal:

\begin{equation}
    \mathbf{G}_{k} = 
    \begin{bmatrix}
    \mathbf{G}_{k,11} & \bm{0} & \dots & \bm{0}\\
    \bm{0} & \mathbf{G}_{k,22} & \dots & \bm{0}\\
    \vdots & \vdots & \ddots & \vdots\\
    \bm{0} &  \bm{0} & \dots & \mathbf{G}_{k,N_g N_g}\\
    \end{bmatrix},
\end{equation}
where $\mathbf{G}_{k,ii}$ is a $d_i\times d_i$ submatrix corresponding to the $i$th mode group with degeneracy $d_i$ and can be written as 
$$\mathbf{G}_{k,ii} = \mathbf{T}_{k,ii} \bm{\Lambda}_{k,ii} \mathbf{T}^H_{k,ii},$$ 
where $\mathbf{T}_{k,ii}$ is a $d_i\times d_i$ unitary matrix, and $\bm{\Lambda}_{k,ii}$ is a $d_i\times d_i$ diagonal matrix of the GDs of the $i$th mode group. In the presence of strong intra-group coupling, 
\begin{align}
    \E{ \mathbf{G}_{k,ii} } ={}& \E{ \mathbf{T}_{k,ii} \bm{\Lambda}_{k,ii} \mathbf{T}^{H}_{k,ii} } \nonumber \\
    ={}& \frac{1}{d_i} \tr{\bm{\Lambda}_{k,ii}} \mathbf{I}_{d_i} = \tau_{0,i} \mathbf{I}_{d_i},
\end{align}
where $\mathbf{I}_{d_i}$ is the $d_i \times d_i$ identity matrix. 
The expectation is over the random unitary group of $d_i\times d_i$ Haar unitary matrices \cite{collins_integration_2006}.
Therefore, we can write $\E{ \mathbf{G}_{k} }$ as 
\begin{equation}
\label{eq:E-Gk}
    \E{ \mathbf{G}_{k} } = \sum_{i=1}^{N_g} \tau_{0,i} \bm{\Gamma}_{i} = \tautau_0 \cdot\Vec{\bm{\Gamma}},
\end{equation}
where $\Vec{\bm{\Gamma}}$ is an $N_g\times 1$ vector of matrices $\{\bm{\Gamma}_{i}\},i=1,\dots,N_g$. $\bm{\Gamma}_{i}$ is a $D\times D$ diagonal matrix whose entries are $1$ only at the indices corresponding to the modes of the $i$th mode group. The $l,m$ element of $\bm{\Gamma}_{i}$ can be written as
\begin{equation}
    \bm{\Gamma}_{i}[l,m] \doteq \delta_{lm} \mathbbm{1}\left(l\in\mathcal{M}_i\right),
    \label{eq:gammaDef}
\end{equation}
where $\mathbbm{1}(\cdot)$ is the indicator function. An example of $\bm{\Gamma}_{1},\bm{\Gamma}_{2},\bm{\Gamma}_{3}$ for $N_g=3$ mode groups with degeneracies $\mathbf{D}=\diag{2,4,6}$ is given below:

\begin{align*}
    \label{eq:gammaEx}
    \bm{\Gamma}_{1} ={}& 
    \begin{bmatrix}
        \mathbf{I}_{2} & \mathbf{0}_{2\times 4} & \mathbf{0}_{2\times 6} \\
        \mathbf{0}_{4\times 2} & \mathbf{0}_{4\times 4} & \mathbf{0}_{4\times 6} \\
        \mathbf{0}_{6\times 2} & \mathbf{0}_{6\times 4} & \mathbf{0}_{6\times 6} \\
    \end{bmatrix},\\
    \bm{\Gamma}_{2} ={}& 
    \begin{bmatrix}
        \mathbf{0}_{2\times 2} & \mathbf{0}_{2\times 4} & \mathbf{0}_{2\times 6} \\
        \mathbf{0}_{4\times 2} & \mathbf{I}_4 & \mathbf{0}_{4\times 6} \\
        \mathbf{0}_{6\times 2} & \mathbf{0}_{6\times 4} & \mathbf{0}_{6\times 6} \\
    \end{bmatrix},\\
    \bm{\Gamma}_{3} ={}& 
    \begin{bmatrix}
        \mathbf{0}_{2\times 2} & \mathbf{0}_{2\times 4} & \mathbf{0}_{2\times 6} \\
        \mathbf{0}_{4\times 2} & \mathbf{0}_{4\times 4} & \mathbf{0}_{4\times 6} \\
        \mathbf{0}_{6\times 2} & \mathbf{0}_{6\times 4} & \mathbf{I}_{6} \\
    \end{bmatrix}.
\end{align*}

The $\{\bm{\Gamma}_{i}\}$ matrices have some important properties:
\begin{subequations}
    \begin{equation}
       \bm{\Gamma}^H_{i} = \bm{\Gamma}^T_{i}=\bm{\Gamma}_{i}.
    \end{equation}
    \begin{equation}
        \tr{\bm{\Gamma}_{i}} = d_i.
    \end{equation}
    \begin{equation}
        \bm{\Gamma}_{i}^n = \bm{\Gamma}_{i}.
    \end{equation}
    \begin{equation}
        \tr{\bm{\Gamma}_{i}^H\bm{\Gamma}_{j}} = \delta_{ij}d_i.
    \end{equation}
\end{subequations}

Owing to strong intra-group coupling, we can represent $\E{ \mathbf{G}_{k} }$ in an $N_g$-dimensional subspace with the basis set  $\{\bm{\Gamma}_{i}\},i=1,\dots,N_g$. 

\subsection*{Analytical Expression for $\E{\mathbf{M}_{k} \mathbf{G}_{k} \mathbf{M}^H_{k}}$}
From the definition of $\mathbf{G}_{k}$, we get
\begin{equation}
\label{eq:E-MkGkMk}
    \mathbf{M}_{k} \mathbf{G}_{k} \mathbf{M}^H_{k} = -j \frac{\partial\mathbf{M}_k}{\partial\omega} \mathbf{M}^H_{k},
\end{equation}
which is the GDO for the backward-propagating signal with transfer function $\mathbf{M}^H_{k}$. The fiber is statistically identical under forward or backward propagation. Therefore, 
\begin{equation}
    \E{\mathbf{M}_{k} \mathbf{G}_{k} \mathbf{M}^H_{k}} = \E{\mathbf{G}_{k} } = \sum_{i=1}^{N_g} \tau_{0,i} \bm{\Gamma}_{i} = \tautau_0\cdot\Vec{\bm{\Gamma}}.
\end{equation}

\subsection*{Analytical Expression for $\tr{ \E{\mathbf{G}^H_{k} \mathbf{G}_{k}} }$ }
We can write the trace of $\E{\mathbf{G}^H_{k} \mathbf{G}_{k}}$ as the sum of the traces of its block-diagonal components: 
\begin{align}
    \begin{split}
        \tr{ \E{\mathbf{G}^H_{k} \mathbf{G}_{k}} } ={}& \sum_{i=1}^{N_g} \tr{\E{\mathbf{G}_{k,ii}^{H} \mathbf{G}_{k,ii}}}
    \end{split}\nonumber \\
    ={}& \sum_{i=1}^{N_g} \tr{\E{ \mathbf{T}_{k,ii} \bm{\Lambda}^2_{k,ii} \mathbf{T}^{H}_{k,ii} }} \nonumber \\
    \begin{split}
                                ={}& \sum_{i=1}^{N_g} \left( d_i \tau_{0,i}^2 + \sigma_{\text{intra},i}^2 \right)
    \end{split} \\
    \begin{split}\label{eq:spanG}
        ={}& \tautau_0^T \mathbf{D} \tautau_0 + \sum_{i=1}^{N_g} \left( \sigma_{\text{intra},i}^2 \right).
    \end{split}
\end{align}
The first term in \eqref{eq:spanG} is the group-delay spread contribution from the average modal delays and the second term is the contribution from the strong intra-group coupling.

\subsection*{Analytical Expression for $\tr{ \E{\mathbf{G}^H_{2} \R \mathbf{M}_{1} \mathbf{G}_{1} \mathbf{M}^H_{1} \R^{H} } }$ }
Using the independence of the fiber spans and the results in \eqref{eq:E-Gk} and \eqref{eq:E-MkGkMk}, we obtain 
\begin{align}
    &\tr{ \E{\mathbf{G}^H_{2} \R \mathbf{M}_{1} \mathbf{G}_{1} \mathbf{M}^H_{1} \R^{H} } } \nonumber \\
    &\qquad= \tr{\E{\mathbf{G}^H_{2}}\R \E{\mathbf{G}_{1}} \R^{H}} \nonumber \\
    &\qquad= \tr{\E{\mathbf{G}_{1}}\R \E{\mathbf{G}_{1}} \R^{H}} \nonumber \\
    &\qquad= \sum_{i=1}^{N_g}\sum_{j=1}^{N_g}\tau_{0,i}\tau_{0,j}\tr{\bm{\Gamma}_{i} \R \bm{\Gamma}_{j} \R^H} \nonumber \\
    &\qquad= \tautau_0^T \PC \tautau_0, \label{eq:GRMGMR}
\end{align}
where $\PC$ is the mode-group power coupling matrix whose $i,j$ element is defined as
\begin{align}
    \PC[i,j] ={}& \tr{\bm{\Gamma}_{i} \R \bm{\Gamma}_{j} \R^H} \nonumber\\
    ={}& \sum_{l\in\mathcal{M}_i}{\sum_{m \in\mathcal{M}_j}{ \left| \R[l,m] \right|^2}}.
    \label{eq:PC_altDef}
\end{align}

\subsection*{Analytical Expression for \\ $\tr{ \E{\mathbf{G}^H_{3} \R \mathbf{M}_{2} \R \mathbf{M}_{1} \mathbf{G}_{1} \mathbf{M}^H_{1} \R^{H} \mathbf{M}^H_{2} \R^{H} } }$ }
Using the independence of the fiber spans and the results in \eqref{eq:E-Gk}, \eqref{eq:E-MkGkMk}, and \eqref{eq:GRMGMR}, we obtain
\begin{align}
    &\trspl{\left( \E{\mathbf{G}^H_{3} \R \mathbf{M}_{2} \R \mathbf{M}_{1} \mathbf{G}_{1} \mathbf{M}^H_{1} \R^{H} \mathbf{M}^H_{2} \R^{H} } \right)} \nonumber \\
    &= \trspl\biggl( \E{\mathbf{G}^H_{3}} \R \nonumber\\
    & \quad\times\E{\mathbf{M}_{2} \R \E{\mathbf{M}_{1} \mathbf{G}_{1} \mathbf{M}^H_{1}} \R^{H} \mathbf{M}^H_{2}} \R^{H} \biggr) \nonumber \\
    \begin{split}
        &= \trspl\biggl( \E{\mathbf{G}^H_{3}} \R \E{\mathbf{M}_{2} \R \E{\mathbf{G}_{1}} \R^{H} \mathbf{M}^H_{2}} \R^{H} \biggr),
    \end{split}
     \label{eq:GRMRGRMR_baby}
\end{align}
\begin{align}
    &\E{\mathbf{M}_{2} \R \E{\mathbf{G}_{1}} \R^{H} \mathbf{M}^H_{2}} \nonumber \\
    &= \E{\mathbf{M}_{2} \R \left(\sum_{i=1}^{N_g} \tau_{0,i} \bm{\Gamma}_{i}  \right) \R^{H} \mathbf{M}^H_{2}} \nonumber \\
    &= \sum_{i=1}^{N_g} \tau_{0,i} \E{\mathbf{M}_{2} \R \bm{\Gamma}_{i} \R^{H} \mathbf{M}^H_{2}} \nonumber \\
    &= \sum_{i=1}^{N_g}{\tau_{0,i} \sum_{j=1}^{N_g}{ \frac{1}{d_j} \tr{\left( \R \bm{\Gamma}_{i} \R^{H} \right)_{jj}} }  \bm{\Gamma}_{j}} \nonumber \\
    &= \sum_{i=1}^{N_g}{\sum_{j=1}^{N_g}{\frac{1}{d_j} \PC[j,i] \tau_{0,i} \bm{\Gamma}_{j}}} = \left(\DPC \tautau_0 \right)\cdot\Vec{\bm{\Gamma}}.\label{eq:MRGRM}
\end{align}

Using \eqref{eq:GRMRGRMR_baby} and \eqref{eq:MRGRM}, we obtain
\begin{align}
    &\tr{ \E{\mathbf{G}^H_{3} \R \mathbf{M}_{2} \R \mathbf{M}_{1} \mathbf{G}_{1} \mathbf{M}^H_{1} \R^{H} \mathbf{M}^H_{2} \R^{H} } } \nonumber \\
    &= \tr{\sum_{i=1}^{N_g}{\sum_{j=1}^{N_g}{\frac{1}{d_j} \PC[j,i] \tau_{0,i} \E{\mathbf{G}^H_{3}}\R\bm{\Gamma}_{j}\R^{H}}}}\nonumber\\
    &= \sum_{i=1}^{N_g}{\sum_{j=1}^{N_g}{\frac{1}{d_j} \PC[j,i] \tau_{0,i} \,\tr{\E{\mathbf{G}^H_{3}}\R\bm{\Gamma}_{j}\R^{H}}}} \nonumber\\
    &= \sum_{i=1}^{N_g}{\sum_{j=1}^{N_g}{\frac{1}{d_j} \PC[j,i] \tau_{0,i} \,\tr{\E{\mathbf{G}^H_{3}}\R\bm{\Gamma}_{j}\R^{H}}}} \nonumber\\
    &= \sum_{i=1}^{N_g}{\sum_{j=1}^{N_g}{\sum_{l=1}^{N_g}{\frac{1}{d_j} \PC[j,i] \tau_{0,i} \tau_{0,l}\PC[l,j]}}} \nonumber\\
    &= \tautau_0^T \PC \DPC \tautau_0. \label{eq:GRMRGRMR}
\end{align}

\subsection*{Analytical Expression for $\E{ \norm{\ttot}^2 }(K)$}

From \eqref{eq:spanG}, \eqref{eq:GRMGMR}, and \eqref{eq:GRMRGRMR}, we can deduce a pattern in the terms in \eqref{eq:gd_formula_baby}. For $k>1$, the $k$th term of \eqref{eq:gd_formula_baby} can be written as $2(K-k+1)\tautau_0^T \PC \left( \DPC\right)^{(k-2)}\tautau_0$. Therefore, \eqref{eq:gd_formula_baby} simplifies to 
\begin{align}
    \begin{split}
            &\E{ \norm{\ttot}^2 }(K) \\
            &\quad= K\left(\tautau_0^T \mathbf{D} \tautau_0 + \sum_{i=1}^{N_g} \left( \sigma_{\text{intra},i}^2 \right) \right) \\
            &\qquad+ 2\left(K-1\right) \tautau_0^T \PC \tautau_0 \\
            &\qquad+ 2\left(K-2\right) \tautau_0^T \PC \left( \DPC \right) \tautau_0 \\
            &\qquad+ \dots \\
            &\qquad+ 2\left(K-k+1\right) \tautau_0^T \PC \left( \DPC\right)^{(k-2)}\tautau_0 \\
            &\qquad+ \dots \\
            &\qquad+ 2\tautau_0^T \PC \left( \DPC\right)^{(K-2)}\tautau_0 \\
            &\quad= K\sum_{i=1}^{N_g} \sigma_{\text{intra},i}^2 + K \tautau_0^T \mathbf{D} \tautau_0 \\
    &\qquad+ \tautau_0^T \left(
    2 \sum_{k=1}^{K-1} \left(K-k \right) \mathbf{D}\left(\DPC \right)^{k} \right)\tautau_0.
    \end{split} \label{eq:gd_jones_pattern}
\end{align}

\section{Derivation of MDL STD in the Generalized Jones Representation}
\label{app:MDL_STD}
In this appendix, we derive the approximate analytical expression for $\E{\norm{\gtotv}^2}$ which is defined as,
\begin{equation*}
    \E{\norm{\gtotv}^2}(K) = \E{\tr{\left(\log \mathbf{M}_\textrm{tot} \mathbf{M}^H_\textrm{tot}\right)^2}}.
\end{equation*}
When the overall MDL is low, we can approximate as follows:
\begin{align}
    \E{\norm{\gtotv}^2}(K)  \approx{}& \E{\tr{\left( \mathbf{M}_\textrm{tot} \mathbf{M}^H_\textrm{tot} - \mathbf{I}_D\right)^2}} \nonumber\\
                            \approx{}& 2\E{\tr{ \mathbf{M}_\textrm{tot} \mathbf{M}^H_\textrm{tot} - \mathbf{I}_D}} \nonumber\\
                            ={}& 2 \tr{\E{\mathbf{M}_\textrm{tot} \mathbf{M}^H_\textrm{tot}}} - 2D. \label{eq:gtot2_approx1}
\end{align}

Using the definition of $\mathbf{M}_\textrm{tot}$, we get
\begin{align}
    &\E{ \mathbf{M}_\textrm{tot} \mathbf{M}^H_\textrm{tot}}  \nonumber \\
    \begin{split}
        &= \Espl\biggl\{\mathbf{M}_{K} \R \dots \mathbf{M}_{2}\R
    \mathbf{M}_{1} \mathbf{M}_{1}^H \R^H \mathbf{M}_{2}^H \dots \R^H \mathbf{M}_{K}^H\biggr\}  \\
    \end{split}\nonumber\\
    &= \E{\mathbf{M}_{K} \R\dots \mathbf{M}_{2}\R\R^H \mathbf{M}_{2}^H \dots \R^H \mathbf{M}_{K}^H } \nonumber\\
    \begin{split}
        &= \Espl\biggl\{\mathbf{M}_{K} \R \dots \E{\mathbf{M}_{2}\R\R^H \mathbf{M}_{2}^H} \dots \R^H \mathbf{M}_{K}^H \biggr\}. 
    \end{split}
    \label{eq:Mtot2}
\end{align}

By repeatedly using \eqref{eq:Uni_exptn_block} in \eqref{eq:Mtot2}, we get
\begin{align}
\begin{split}
    &\E{\mathbf{M}_{2}\R\R^H \mathbf{M}_{2}^H} = \sum_{i=1}^{N_g}{ \sum_{j=1}^{N_g}{\frac{1}{d_j} \PC[j,i]\bm{\Gamma}_i}} \\
    &\qquad= \sum_{i=1}^{N_g}\left( \DPC\mathbf{1}_{N_g}\right)_i\bm{\Gamma}_i,
\end{split}\\
\begin{split}
    &\E{\mathbf{M}_{3}\E{\mathbf{M}_{2}\R\R^H \mathbf{M}_{2}^H} \mathbf{M}_{3}^H} \\
    &\qquad= \sum_{i=1}^{N_g}\left(\left( \DPC\right)^2\mathbf{1}_{N_g}\right)_i\bm{\Gamma}_i,
\end{split}\\
&\qquad\vdots\nonumber\\ 
\begin{split}
    &\E{\mathbf{M}_{K} \R \dots \E{\mathbf{M}_{2}\R\R^H \mathbf{M}_{2}^H} \dots \R^H \mathbf{M}_{K}^H } \\
    &\qquad=\sum_{i=1}^{N_g}\left(\left( \DPC\right)^{K-1}\mathbf{1}_{N_g}\right)_i\bm{\Gamma}_i.
\end{split}\label{eq:Mtot2_all}
\end{align}

 Using \eqref{eq:Mtot2_all} in \eqref{eq:gtot2_approx1} we get
\begin{equation}
    \E{\norm{\gtotv}^2}(K)  \approx 2\left(\mathbf{d}^T \left( \DPC\right)^{K-1}\mathbf{1}_{N_g}-D \right).
\end{equation}



\section{Modeling Mode Scramblers in the Generalized Stokes Space}
\label{app:Stokes_Model}
Assuming no MDL, the evolution of the modal dispersion along the fiber link can be modeled via stochastic differential equations in the generalized Stokes space by \cite{antonelli_stokes-space_2012, arik_delay_2015}
\begin{align}
    \frac{\partial\tauS}{\partial z} &= \dBdw + \betazs \times \tauS \label{eq:stokes_stoch_tau},\\
    \betazs(z) &= \dBdw + \mathbf{w}(z) \label{eq:stokes_stoch_biref}
\end{align}
where $\tauS$ is the $(D^2 - 1) \times 1$ dispersion vector representing the GDO in the generalized Stokes space, $\betazs$ is the $(D^2 - 1) \times 1$ birefringence vector, $\dBdw = \partial \betazs / \partial \omega$ is a $(D^2 - 1) \times 1$ vector representing the deterministic GD of the GI-MMF fiber, $z$ is the propagation distance, and $\mathbf{w}(z)$ is the $(D^2 - 1) \times 1$ additive noise vector representing mode coupling. 
Conversion from the generalized Jones representation to the generalized Stokes representation and the cross product is explained in Appendix \ref{app:Stokes_Space_Rep}.


Following \cite{arik_delay_2015}, we assume $\mathbf{w}(z)$ is distributed as $\mathbf{w}(z) \sim \mathcal{N}(\bm{0},\HH)$ where $\HH$ is a positive-definite real-valued covariance matrix, and $\mathbf{W}(l) = \int_0^l\HH^{0.5}\mathbf{w}(z)dz$ is a $(D^2 - 1) \times 1$-dimensional Wiener process, such that $\mathbf{W}(0)=\bm{0}$ and $\mathbf{W}(z) \sim \mathcal{N}(\bm{0}, z\mathbf{I})$. 


\eqref{eq:stokes_stoch_tau} and \eqref{eq:stokes_stoch_biref} can be combined compactly into
\begin{equation*}
    \partial\tauS = \left(\dBdw + \omega \dBdw  \times \tauS \right)\partial z - \tauS \times \HH^{0.5}\partial\mathbf{W}. \label{eq:stokes_stoch}
\end{equation*}
The moments of $\tauS$ can be described by two deterministic ordinary differential equations (ODE)s \cite{arik_delay_2015}:
\begin{align}
    \ddz\E{\tauS} &= \dBdw - \Q  \E{\tauS}, \label{eq:evo_tau} \\
    \ddz\E{\norm{\tauS}^2} &= 2\dBdw^T\E{\tauS}, \label{eq:evo_tausq}
\end{align}
where $\mathbf{Q}$ is evaluated from $\HH$ and the structure coefficients $f$, defined in Appendix \ref{app:Stokes_Space_Rep} by
\begin{equation*}
    \Q[j,l]=\frac{1}{2}\sum_{m,k=1}^{D^2-1}f[k,m,l]f[k,m,j]\HH[k,k].
\end{equation*}
The GD STD can be obtained as \cite{arik_delay_2015}
\begin{equation*}
    \sigma_{\mathrm{GD}} = \sqrt{\E{\norm{\tauS}^2}/D}.
\end{equation*}

We can include the effect of the mode scrambler $\R$ at a specified propagation distance $z_k$  into the dispersion evolution equations (\ref{eq:evo_tau}) and (\ref{eq:evo_tausq}) by an abrupt rotation of the expected dispersion vector: 
\begin{subequations}
    \begin{align}
    \E{\norm{\tauS\left(z_k^+ \right)}^2} &= \E{\norm{\tauS\left(z_k^- \right)}^2}, \\
    \E{\tauS\left(z_k^+ \right)} &= \RS\E{\tauS\left(z_k^- \right)},
    \end{align} \label{eq:ms_bc}
\end{subequations} 
where $\RS$ is the mode scrambler matrix in generalized Stokes space given by
\begin{equation*}
    \RS[i,j]=\tr{\bm{\Lambda}_j\R^H\bm{\Lambda}_i\R}.
\end{equation*}
We assume the mode scrambler does not add any modal dispersion and has negligible MDL.


We integrate \eqref{eq:evo_tau} and \eqref{eq:evo_tausq} for arbitrary initial conditions,
\begin{subequations}
    \begin{align}
    &\E{\tauS \left(0\right)}=\RS\tauSinit, \\
    &\E{\norm{\tauS\left(0\right)}^2}=\sigma_{\textrm{init}}^2,
    \end{align} \label{eq:mode_scrambler_bc}
\end{subequations}
yielding
\begin{equation}
    \begin{split}
        \E{\tauS} &= \Q^{-1}\left(\mathbf{I} - \exp{(-z\Q)} \right) \dBdw  \\ &+ \exp{(-z\Q)}\RS\tauSinit, \label{eq:gen_tau_integral}
    \end{split}
\end{equation}
\begin{align}
    \begin{split}
        \E{\norm{\tauS}^2} &= 2 \dBdw^T \left(\Q^{-1}z - \Q^{-2}\left(\mathbf{I} - \exp{(-z\Q)} \right) \right)\dBdw  \\
        & + 2 \dBdw^T\left(\mathbf{I} - \exp{(-z\Q)} \right)\RS\tauSinit + \sigma_{\textrm{init}}^2.\label{eq:gen_tausq_integral}
    \end{split}
\end{align}
For a system with multiple spans and mode scramblers after every span, the GD STD of the $k$-th span can be obtained by repeated application of the boundary conditions \eqref{eq:mode_scrambler_bc} and integration of \eqref{eq:evo_tau} and \eqref{eq:evo_tausq} with the appropriate initial conditions.
To obtain a closed-form expression, we leverage the structure of $\Q$ and $\dBdw$ to simplify \eqref{eq:gen_tau_integral} and \eqref{eq:gen_tausq_integral} substantially. 
We directly evaluate for a GI-MMF with $D = 12$ modes and $N_g = 3$ mode groups and find that only a small subset of the $D^2-1$ elements of $\tauS$ are needed to accurately compute $\sigma_{\mathrm{GD}}$.

\subsection{Direct Evaluation for \texorpdfstring{$D = 12$}{D=12} Modes and \texorpdfstring{$N_g = 3$}{Ng = 3} Mode Groups}

In deriving the analytic formula of GD STD, we use the same indexing of modes as in Section II and make the following assumptions to simplify the analysis:
\begin{enumerate}[label=(\alph*)]
    \item Modes within the same mode group strongly couple to each other with intra-group coupling length $\lenT$. 
    \item Modes not in the same mode group weakly couple to each other with inter-group coupling length $\lenO$.
    \item There is negligible polarization mode dispersion.
    \item The only source of intra-group delay spread comes from the third mode group as its spatial modes $\mathrm{LP}_{\mathrm{02}}$ and $\mathrm{LP}_{\mathrm{21}}$ are not exactly degenerate. 
\end{enumerate}
Under these assumptions, the uncoupled GD per unit lengths will satisfy the following:
\begin{align*}
        \beta_{1,1} &= \beta_{1,2},  \\
        \beta_{1,3} = \beta_{1,4} &= \beta_{1,5} = \beta_{1,6},  \\
        \beta_{1,7} = \beta_{1,8} &= \beta_{1,9} = \beta_{1,10}, \\
        \beta_{1,11} &= \beta_{1,12}.\\
\end{align*}
We assign the following uncoupled inter- and intra-group delay differences:
\begin{align*}
        \DelBeta{1 \mathrm{-} 2}&=\beta_{1,1} - \beta_{1,3},  \\ 
        \DelBeta{2\mathrm{-}3}&=\beta_{1,3} - \beta_{1,7}, \\ 
        \DelBeta{3\mathrm{-}3} &= \beta_{1,7} - \beta_{1,11}, 
\end{align*}
where $\DelBeta{1 \mathrm{-} 2}$ is the uncoupled inter-group delay difference between mode groups $1$ and $2$, $\DelBeta{2\mathrm{-}3}$ is the uncoupled inter-group delay difference between mode groups $2$ and $3$, and $\DelBeta{3\mathrm{-}3}$ is the uncoupled intra-group delay difference between the spatial modes $\mathrm{LP}_{02}$ and $\mathrm{LP}_{21}$ of mode group $3$.

Following the above-mentioned assumptions, $\HH$ is diagonal with diagonal elements of either $2L^{-1}_{\textrm{inter}}$ or $2L^{-1}_{\textrm{intra}}$. 
Using the generalized Pauli matrix basis $\tilde{\mathbf{\Gamma}}_i$ $(1\leq i\leq D^2-1)$ defined in the Appendix of \cite{antonelli_stokes-space_2012}, for $D=12$ with mode-group degeneracy $\mathbf{d}=[2,4,6]^T$,
\begin{align*}
    \begin{split}
    &\HH[k,k] = \\ &\begin{cases}
			2/L_{\textrm{intra}}, & k \in [1, 18] \cup [27, 30] \cup [55, 58] \cup [71, 143] \\
            2/L_{\textrm{inter}}, & k \in [19, 26] \cup [31, 54] \cup [59,70].
		 \end{cases} 
    \end{split} 
\end{align*}
Direct evaluation of $\Q$ yields 
\begin{align*}
\begin{split}
    \Q &= \begin{bmatrix}
         \Qone & \mathbf{0}_{(D^2-\frac{D}{2})\times(\frac{D}{2}-1)} \\
         \mathbf{0}_{(\frac{D}{2}-1)\times(D^2-\frac{D}{2})} & \Qsub
        \end{bmatrix}  \\ &=
    \begin{bmatrix}
         \Qone & \mathbf{0}_{138\times5} \\
         \mathbf{0}_{5\times138} & \Qsub
        \end{bmatrix}. 
\end{split}
\end{align*}
We obtain a larger block $\Qone$ with dimension $(D^2-D/2) \times (D^2-D/2)$ and a smaller block $\Qsub$ with dimension $(D/2-1) \times (D/2-1)$.
$\Qone$ is diagonal with elements $\Qone[i,i] \propto L_{\textrm{intra}}^{-1}$ and $\Qsub$ is real and symmetric.

\begin{figure*}[ht!]
    \hrulefill
    
    \begin{equation}
        \begin{split}
         &\mathbf{V}_{\mathrm{\RomanNumeralCaps{2}}} = \begin{bmatrix}
            0 & 0 & -1/2 & \sqrt{3}/2 & 0 \\
            0 & 0 & \sqrt{3}/2 & 1/2 & 0 \\
            -\sqrt{2}/4 & \sqrt{3/8} & 0 & 0 & \sqrt{2}/2 \\
            -\sqrt{30}/20 & -\sqrt{5/8} & 0 & 0 & \sqrt{30}/10 \\
            2\sqrt{5}/5 & 0 & 0 & 0 & \sqrt{5}/5 \\
            \end{bmatrix} = [\vtil_1, \: \vtil_2, \: \vtil_3, \: \vtil_4, \: \vtil_5]
        \end{split} \label{eq:V_Q}
    \end{equation}
     \begin{equation}
        \begin{split}
        &\mathbf{D}_{\mathrm{\RomanNumeralCaps{2}}} = \begin{bmatrix}
            L_{\textrm{intra}}^{-1} + L_{\textrm{inter}}^{-1} & 0 & 0 & 0 & 0 \\
            0 & L_{\textrm{intra}}^{-1} + L_{\textrm{inter}}^{-1} & 0 & 0 & 0 \\
            0 & 0 & \frac{2}{3}L_{\textrm{intra}}^{-1}+\frac{4}{3}L_{\textrm{inter}}^{-1} & 0 & 0 \\
            0 & 0 & 0 & 2L_{\textrm{inter}}^{-1} & 0 \\
            0 & 0 & 0 & 0 & 2L_{\textrm{inter}}^{-1} \\
            \end{bmatrix}
        \end{split} \label{eq:D_Q}
    \end{equation}   
    
    \hrulefill
    \label{fig:D_12_submatrices}
    
\end{figure*}

Eigenvalue decomposition of $\Qsub$ yields $\Qsub = \mathbf{V}_{\mathrm{\RomanNumeralCaps{2}}}\mathbf{D}_{\mathrm{\RomanNumeralCaps{2}}}\mathbf{V}^{T}_{\mathrm{\RomanNumeralCaps{2}}}$. $\mathbf{V}_{\mathrm{\RomanNumeralCaps{2}}}$ and $\mathbf{D}_{\mathrm{\RomanNumeralCaps{2}}}$ are shown in equations \eqref{eq:V_Q} and \eqref{eq:D_Q}.
We see that $\mathbf{D}_{\mathrm{\RomanNumeralCaps{2}}}$ contains two eigenvalues equal to $2L_{\textrm{inter}}^{-1}$ and three composed of linear combinations of $L_{\textrm{intra}}^{-1}$ and $L_{\textrm{inter}}^{-1}$.
Similarly, direct evaluation of $\dBdw$ yields
\begin{equation*}
    \dBdw = \begin{bmatrix}
         \mathbf{0}_{138\times1}  \\
         \dBdwsub
        \end{bmatrix}, \label{eq:beta_block}
\end{equation*}
where 
\begin{equation*}
    \begin{split}
        \dBdwsub &= 
            2\left(\DelBeta{1 \mathrm{-} 2} + 3\DelBeta{2\mathrm{-}3} + \DelBeta{3\mathrm{-}3}\right)\vtil_5 \\
            &+ 4\DelBeta{1 \mathrm{-} 2}\vtil_4 + 4\DelBeta{3\mathrm{-}3}\vtil_1. 
    \end{split} 
\end{equation*}   

Similarly to $\Q$, $\dBdw$ can be partitioned into two parts.
Furthermore, $\dBdwsub$ can be decomposed into three eigenvectors of $\Qsub$: $\vtil_1$, $\vtil_4$, and $\vtil_5$. 
Eigenvectors $\vtil_4$ and $\vtil_5$ share the same eigenvalue $2L_{\textrm{inter}}^{-1}$, which is related to the inter-group coupling length.
Eigenvector $\vtil_1$ has eigenvalue $L_{\textrm{intra}}^{-1} + L_{\textrm{inter}}^{-1}$, which will be dominated by $L_{\textrm{intra}}$ since $L_{\textrm{intra}} \ll L_{\textrm{inter}}$ in typical GI-MMF fibers.
Therefore, we define 
\begin{subequations}
    \begin{equation}
    \begin{split}
        \dBinter &= 2\left(\DelBeta{1 \mathrm{-} 2} + 3\DelBeta{2\mathrm{-}3} + \DelBeta{3\mathrm{-}3}\right)\vtil_5 \\ &+ 4\DelBeta{1 \mathrm{-} 2}\vtil_4 \label{eq:dBdw_decomp1}
    \end{split}
    \end{equation}
    \begin{equation}
        \dBintra = 4\DelBeta{3 \mathrm{-} 3}\vtil_1 \qquad \qquad \qquad \qquad \qquad \qquad \label{eq:dBdw_decomp2}
    \end{equation}
\end{subequations}
so we can decompose $\dBdwsub$ as
\begin{equation*}
    \dBdwsub = \dBinter + \dBintra. \
\end{equation*}
We note that  $\vtil_2^T\dBdwsub$ and $\vtil_3^T\dBdwsub$ extract the uncoupled GD differences between $\mathrm{LP}_{21a}$ and $\mathrm{LP}_{21b}$, and between $\mathrm{LP}_{11a}$ and $\mathrm{LP}_{11b}$, respectively. Owing to our assumptions on the source of intra-group delay spread, the uncoupled GD differences between these modes are 0.

Through direct evaluation, we observe a reduction in the dimensionality of the terms in \eqref{eq:gen_tau_integral} and \eqref{eq:gen_tausq_integral} since $\dBdw$ is non-zero in only the last $D/2-1$ elements.
Only the last term in \eqref{eq:gen_tau_integral}, namely, $\exp{(-z\Q)}\RS\tauSinit$, is non-zero in all $D^2-1$ elements, since $\RS$ is generally non-zero in all $(D^2-1) \times (D^2-1)$ elements.
However, the first $D^2-D/2$ elements of $\RS\tauSinit$ will decay exponentially with the intra-group coupling length since $\Qone[i,i] \propto L_{\textrm{intra}}^{-1}$.
Hence to the zeroth order, the evolution of $\tauS$ can be captured by its last $D/2-1$ elements expressed by $\tauSsub$.
Rewriting \eqref{eq:gen_tau_integral} and \eqref{eq:gen_tausq_integral} using this simplification gives
\begin{align*}
    \begin{split}
            \E{\tauSsub} &= \Qsub^{-1}\left(\mathbf{I} - \exp{(-z\Qsub)} \right) \dBdwsub \\ &+ \exp{(-z\Qsub)} \RSsubinit \tauSinitsub, 
    \end{split}
\end{align*}
\begin{align*}
    \begin{split}       
        &\E{\norm{\tauS}^2}  \\ 
        &=  2 \dBdwsub^T \left(\Qsub^{-1}z - \Qsub^{-2}\left(\mathbf{I} - \exp{(-z\Qsub)} \right) \right)\dBdwsub  \\
        & + 2 \dBdwsub^T\left(\mathbf{I} - \exp{(-z\Qsub)} \right) \RSsubinit \tauSinitsub + \sigma_{\textrm{init}}^2, 
    \end{split} 
\end{align*}
where we have used $\E{\norm{\bm{\tau}}^2} \approx \E{\norm{\tauSsub}^2}$ and $\RSsubinit$ is the lower $(D/2-1) \times (D/2-1)$ block of $\RS$. 
We define $\zone{z}{L}$ and $\ztwo{z}{L}$,
\begin{subequations}
    \begin{equation*}
    \zone{z}{L} = L\left(1-\exp{\left(-\frac{z}{L}\right)} \right),
    \end{equation*}
    \begin{equation*}
    \ztwo{z}{L} = Lz-L^2\left(1-\exp{\left(-\frac{z}{L}\right)} \right).
    \end{equation*}
\end{subequations}
Using \eqref{eq:dBdw_decomp1} and \eqref{eq:dBdw_decomp2} we obtain
\begin{align}
    \begin{split}
        \E{\tauSsub} &= \zone{z}{\lenOH}\dBO \\ 
        &+ \zone{z}{\lenTH}\dBT  \\ 
        &+ \exp{(-z\Qsub)} \RSsubinit \tauSinitsub, \label{eq:sub_tau_integral_simple} 
    \end{split}
\end{align}
and
\begin{align}
    \begin{split}       
        \E{\norm{\tauS}^2} &=  2 \ztwo{z}{\lenOH}\norm{\dBO}^2 \\ 
        &+ 2 \ztwo{z}{\lenTH}\norm{\dBT}^2 \\
        &+ 2\zone{z}{\lenOH}\dBO^T\RSsubinit \tauSinitsub \\ 
        &+ 2\zone{z}{\lenTH}\dBT^T \RSsubinit \tauSinitsub \\
        &+ \sigma_{\textrm{init}}^2, \label{eq:sub_tausq_integral_simple} 
    \end{split}
\end{align}
where $2L_{\textrm{mix}}^{-1} = L_{\textrm{intra}}^{-1} + L_{\textrm{inter}}^{-1}$.

Employing the simplifications in \eqref{eq:sub_tau_integral_simple} and \eqref{eq:sub_tausq_integral_simple}, equations \eqref{eq:stokes_stoch_tau} and \eqref{eq:stokes_stoch_biref} can be solved for the system described in Section II. The initial conditions are $\E{\bm{\tau}\left(0\right)} = 0$ and $\E{\norm{\tautau\left(0\right)}^2} = 0$. The following piecewise analytic expression is obtained. For $z \in [K\lenS, (K+1)\lenS)$,
\begin{align}
    \begin{split}
         &\E{\norm{\tauS \left(z\right)}^2} \\ 
         &= 2 \ztwo{z-K\lenS}{\lenOH}\norm{\dBO}^2 \\
         &+ 2 \ztwo{z-K\lenS}{\lenTH}\norm{\dBT}^2 \\
         &+ 2 \zone{z-K\lenS}{\lenOH}\dBO^T \RSsub\left(\sum_{k=1}^{K-1}\left(\PCS\right)^k\right)\tauSZ \\
         &+ 2 \zone{z-K\lenS}{\lenTH}\dBT^T \RSsub\left(\sum_{k=1}^{K-1}\left(\PCS\right)^k\right)\tauSZ \\
         &+ K \sigSZ + 2\tauSZ^T\left(\sum_{k=1}^{K-1}(K-k)\DSinv\left(\PCS\right)^k\right)\tauSZ \label{eq:stokes_gds}
    \end{split}
\end{align}
where 
\begin{align*}
    \begin{split}
        \tauSZ &= \zone{\lenS}{\lenOH}\dBO \\
            &+ \zone{\lenS}{\lenTH}\dBT, 
    \end{split}
\end{align*}
\begin{align*}
    \begin{split}
      \sigSZ &= 2 \ztwo{\lenS}{\lenOH}\norm{\dBO}^2 \\
       &+ 2 \ztwo{\lenS}{\lenTH}\norm{\dBT}^2, 
    \end{split}
\end{align*}
\begin{align*}
    \DS = \exp{ \left( -\lenS \Qsub \right)},
\end{align*}
\begin{equation*}
    \RSsub = \RSsubinit.
\end{equation*}

$\tauSZ$ and $\sigSZ$ are the expected dimensionally reduced dispersion vector and norm-square of the dispersion vector after a span length of accumulation with no mode scrambling, respectively. $\RSsub$ is the analogous quantity to the power coupling matrix $\PC$ in generalized Stokes space. $\DS$ captures the distortion of the Stokes power coupling matrix by finite coupling lengths after a span length.

\subsection{Comparison to the Generalized Jones Space}
Comparing \eqref{eq:stokes_gds} evaluated at $z=K\lenS$ with the Jones formula reveals the following connections:
\begin{subequations}
    \begin{align*}
        \sigSZ &\longleftrightarrow \sum_{i=1}^{N_g} \sigma_{\text{intra},i}^2 + \tautau_0^T \mathbf{D} \tautau_0, \\
        \tauSZ &\longleftrightarrow \tautau_0, \\
        \RSsub &\longleftrightarrow \PC, \\
        \PCS   &\longleftrightarrow \DPC.
    \end{align*}
\end{subequations}
The main difference between the two formulae is the inclusion of the exponentially decaying terms arising from the finite intra- and inter-group coupling lengths and that $\tauSZ$ also contains contributions from the intra-group delay spread. We can obtain the Jones formula from \eqref{eq:stokes_gds}, up to a scale factor of $\sqrt{D}$, by ignoring the effect of mode scrambling on the intra-group delay spread accumulation,
\begin{subequations}
    \begin{align*}
        \zone{z-K\lenS}{\lenTH}\dBT^T \RSsub\left(\sum_{k=1}^{K-1}\left(\PCS\right)^k\right)\tauSZ  \approx 0,
    \end{align*}
\end{subequations}
and making the following approximations: 
\begin{subequations}
    \begin{align*}
        \zone{\lenS}{\lenOH} &\approx \lenS, \\
        \ztwo{\lenS}{\lenOH} &\approx \frac{\lenS^2}{2}, \\
        \DS &\approx \vtil_4\vtil_4^T + \vtil_5\vtil_5^T.
    \end{align*}
\end{subequations}

\section{Operator Representations in the Generalized Stokes Space}
\label{app:Stokes_Space_Rep}
In the generalized Stokes space, the GD operator can be represented in the basis of generalized Pauli matrices $\tilde{\mathbf{\Gamma}}_i$ $(1\leq i\leq D^2-1)$. 
We use the algorithm in Appendix A of \cite{antonelli_stokes-space_2012} to generate the generalized Pauli matrices.
The $(D^2-1)$-dimensional vector $\tauS = [\tau_1 \ldots \tau_{D^2-1}]^T$ is the generalized Stokes space vector corresponding to $\mathbf{G}_{\textrm{tot}}$, where $\tauS_i = \tr{\tilde{\mathbf{\Gamma}}_i \mathbf{G}_{\textrm{tot}}}$. 
Similarly, for a fiber with propagation constants $\beta_{0,1}, \dots, \beta_{0,D}$ and uncoupled group delays $\beta_{1,1}, \dots, \beta_{1,D}$, the generalized Stokes space vectors for the propagation constants $\betazs$ and the group delays $\dBdw$ can be evaluated from 
\begin{subequations}
    \begin{align*}
        \betazsi &= \tr{\tilde{\mathbf{\Gamma}}_i \diag{\beta_{0,1}, \cdots, \beta_{0,D}}}, \\
        \dBdwi &= \tr{\tilde{\mathbf{\Gamma}}_i \diag{\beta_{1,1}, \cdots, \beta_{1,D}}}.
    \end{align*}
\end{subequations}
$\dBdw$ of a fiber with group delays $\diag{\left(\beta_{1,1}, \cdots, \beta_{1,D}\right)}$ can be found using  $\dBdw = \tr{\diag{\tau_1, \cdots, \tau_D}}$
The dot product operations between two $D^2 - 1$-dimensional vectors $\mathbf{a}$ and $\mathbf{b}$ in the generalized Stokes Space is 
\begin{equation*}
    \mathbf{a}\cdot\mathbf{b} = \sum_{i=1}^{D^2-1}a_ib_i.
\end{equation*}
The cross product between $\mathbf{a}$ and $\mathbf{b}$ is
\begin{equation*}
    \mathbf{a}\times\mathbf{b} = \sum_{i,j,k=1}^{D^2-1}f[i,j,k]a_ib_j\mathbf{e}_k,
\end{equation*}
where $\mathbf{e}_k$ are the orthogonal unit-length basis vectors of the generalized Stokes space and $f[i,j,k]$ is a structure coefficient given by
\begin{equation*}
    f[i,j,k] = \frac{\jmath}{D^2}\tr{\tilde{\mathbf{\Gamma}}_k\left(\tilde{\mathbf{\Gamma}}_i\tilde{\mathbf{\Gamma}}_j - \tilde{\mathbf{\Gamma}}_j\tilde{\mathbf{\Gamma}}_i \right)}.
\end{equation*}
A more detailed review of Stokes space operator representations can be found in Appendix A of \cite{arik_delay_2015}.
\section{Derivation of GD STD in the Generalized Stokes Representation}
\label{app:RMS_GDS_Stokes}
In this appendix, we provide the derivation of \eqref{eq:stokes_gds}. We repeatedly use \eqref{eq:ms_bc} to obtain the initial conditions at every new span and use \eqref{eq:sub_tau_integral_simple} and \eqref{eq:sub_tausq_integral_simple} to solve for $\E{\tauS}$ and $\E{\norm{\tauS}^2}$ in each span. We assume the initial modal dispersion and GD STD are 0. We evaluate the GD STD accumulation for the first three spans and notice the same pattern of terms as in \eqref{eq:gd_jones_pattern}.

For $z \in [0, \lenS)$ the initial conditions are 
\begin{align*}
&\E{\tauS \left(0\right)}=\mathbf{0} \\
&\E{\norm{\tauS\left(0\right)}^2}=0.
\end{align*}
The solution of the evolution equations is:
\begin{align*}
        \begin{split}
            \E{\tauSsub} &= \zone{z}{\lenOH}\dBO \\
            &+ \zone{z}{\lenTH}\dBT 
        \end{split}
\end{align*}

\begin{align*}
    \begin{split}
        \E{\norm{\tauS}^2} &= 2 \ztwo{z}{\lenOH}\norm{\dBO}^2 \\
         &+ 2 \ztwo{z}{\lenTH}\norm{\dBT}^2.
    \end{split}
\end{align*}
For $z \in [\lenS, 2\lenS)$ the boundary conditions are 
\begin{align*}
    \begin{split}
        \E{\tauS \left(\lenS^+\right)} = \RSsub \tauSZ &= \RSsub \zone{\lenS}{\lenOH}\dBO \\
            &+ \RSsub\zone{\lenS}{\lenTH}\dBT 
    \end{split}
\end{align*}

\begin{align*}
    \begin{split}
       \E{\norm{\tauS\left(\lenS^+\right)}^2}= \sigSZ &= 2 \ztwo{\lenS}{\lenOH}\norm{\dBO}^2 \\
       &+ 2 \ztwo{\lenS}{\lenTH}\norm{\dBT}^2.
    \end{split}
\end{align*}
The solution of the dispersion evolution equations is
\begin{align*}
    \begin{split}
        \E{\tauSsub} &= \zone{z-\lenS}{\lenOH}\dBO \\
        &+ \zone{z-\lenS}{\lenTH}\dBT  \\
        &+ \Qexp{z-\lenS} \RSsub \tauSZ
    \end{split}
\end{align*}

\begin{align*}
    \begin{split}
        \E{\norm{\tauS}^2} &= 2 \ztwo{z-\lenS}{\lenOH}\norm{\dBO}^2 \\
         &+ 2 \ztwo{z-\lenS}{\lenTH}\norm{\dBT}^2 \\
         &+ 2\zone{z-\lenS}{\lenOH}\dBO^T \RSsub\tauSZ \\
         &+ 2\zone{z-\lenS}{\lenTH}\dBT^T \RSsub\tauSZ + \sigSZ.
    \end{split}
\end{align*}
For $z$ $\in$ $[2\lenS$$,$$ 3\lenS)$, the boundary conditions are
\begin{align*}
    \begin{split}
        \E{\tauS \left(2\lenS^+\right)} = \RSsub\left(1 + \PCS\right)\tauSZ
    \end{split}
\end{align*}
\begin{align*}
    \begin{split}
       \E{\norm{\tauS\left(2\lenS^+\right)}^2}= 2\sigSZ + 2\tauSZ^T \RSsub \tauSZ.
    \end{split}
\end{align*}
The solution of the evolution equations is
\begin{align*}
    \begin{split}
        \E{\tauSsub} &= \zone{z-2\lenS}{\lenOH}\dBO \\
        &+ \zone{z-2\lenS}{\lenTH}\dBT  \\
        &+ \Qexp{z-2\lenS} \RSsub\left(1 + \PCS\right)\tauSZ   
    \end{split}
\end{align*}
\begin{align*}
    \begin{split}
        &\E{\norm{\tauS}^2} \\ 
         &= 2 \ztwo{z-2\lenS}{\lenOH}\norm{\dBO}^2 \\
         &+ 2 \ztwo{z-2\lenS}{\lenTH}\norm{\dBT}^2 \\
         &+ 2\zone{z-2\lenS}{\lenOH}\dBO^T \RSsub\left(1 + \PCS\right)\tauSZ \\
         &+ 2\zone{z-2\lenS}{\lenTH}\dBT^T \RSsub\left(1 + \PCS\right)\tauSZ \\
         &+ 2\sigSZ + 2\tauSZ^T \RSsub \tauSZ.
    \end{split}
\end{align*}

We recover the same pattern of terms as in \eqref{eq:gd_jones_pattern}. Continuing this pattern, we can obtain \eqref{eq:stokes_gds}. 

\section{Mode Scrambling as a Markov Chain}
\label{app:Markov_model}
The theory of Markov chains is useful in the analysis of mode scrambling. The theorems, definitions, and notation in the following analysis are taken from Meyn and Tweedie \cite{meyn_markov_2009}.
We define a Markov chain $\bPhi = \{\bPhi_0, \bPhi_1, \dots\}$ on the evolution of the system transfer matrix for an integer number of spans over the state space $\X$. For $k\geq1$,
\begin{equation}
\label{eq:markov}
    \bPhi_k = \mathbf{R} \mathbf{M}_k \bPhi_{k-1},
\end{equation}
where the initial state can be chosen to be $\bPhi_0=\mathbf{I}_D$, the $D\times D$ identity matrix,
$\{\mathbf{M}_k\}$ are random $\mathbf{d}$-block-unitary matrices with block sizes given by the degeneracy vector $\mathbf{d}$, and  $\mathbf{R}$ is the deterministic transfer matrix of the mode scramblers.
It is easy to see that $\bPhi$ is indeed a time-homogeneous Markov chain \cite{meyn_markov_2009} when $\{\mathbf{M}_k\}$ are independent of each other--the conditional probability distribution of $\bPhi_k$ given $\bPhi_{k-1}$ is independent of $\bPhi_{k-2},...,\bPhi_0$.

The Markov state $\bPhi_k$ is a $D\times D$ continuous, complex-valued matrix.
For the rest of this analysis, we assume the following:
\begin{itemize}
    \item $\{\mathbf{M}_k\}$ are independent and identically distributed (IID) random matrices distributed according to the Haar measure on the group of $\mathbf{d}$-block-unitary matrices $\U_\mathbf{d}$
    \item $\mathbf{R}$ is a unitary matrix satisfying the design criteria for strong mode scrambling in Section III.
    \item The eigenvalues of $\mathbf{R}$ are of the form $\{e^{\jmath \theta_i}, 1\leq i \leq D\}$ such that $\{\theta_i\}$ are rational multiples of $2\pi$.
\end{itemize}
Under the first two assumptions, the state space $\X$ is the compact topological group of $D\times D$ unitary matrices $\U_D$. 
We use the third assumption to prove an important property of the Markov chain in Lemma \ref{lemma:A1_subgroup} and repeatedly use the property in theorems thereafter.
Postulate \ref{pos:A1_subgroup} attempts to remove the dependence of the proof on this condition on $\R$.
However, in the practical case, any $\R \in \U_D$ can be approximated to arbitrary precision to satisfy this condition.

The Markov state is related to the total system transfer matrix after $k$ spans, $\mathbf{M}_\textrm{tot}(k) = \mathbf{R}^H\bPhi_k$. 
We are interested in the distribution of $\mathbf{M}_\textrm{tot}(k)$ and $\mathbf{G}_\textrm{tot}(k)$ for large $k$. 
This is equivalent to finding the distribution of $\bPhi_k$.

In this appendix, we prove that the chain $\bPhi$ is $\psi$-irreducible, aperiodic, positive, Harris recurrent, and uniformly ergodic.
These properties ensure that the Law of Large Numbers (LLN) and the Central Limit Theorem (CLT) hold for a large class of functions on the Markov state, including the GDO.

\subsection*{Transition Probability and Function}
Let $\bX$ denote the Borel $\sigma$-field over $\X$.
The one-step transition probability of the Markov chain $\bPhi$ is defined for every $x\in \X$, and every set $A\in \bX$,
\begin{align}
    \nonumber P(x,A)  &= \prob(\bPhi_1 \in A | \bPhi_0 = x) \\ \label{eq:transn_prob} \nonumber
            &= \prob(\R \mathbf{M} x \in A) \\
            &= \int_{\U_\mathbf{d}} \mu_\mathbf{d}(dm) \mathbbm{1}(\R mx\in A),
\end{align}
where $\mathbf{M}$ is a random $\mathbf{d}$-block-unitary matrix distributed according to the Haar measure.
Similarly, we can write the $k$-step transition probability, $P^k(x,A) = \prob(\bPhi_k \in A | \bPhi_0 = x)$.

The one-step transition function $\f:\U_{D} \times \U_\mathbf{d}\rightarrow\U_{D}$ is defined as
\begin{equation}
\label{eq:transn_func}
    \f(x,m_1) = \R m_1 x.
\end{equation}
Similarly, the $k$-step transition function is the map $\f_k:\U_{D} \times (\U_\mathbf{d} \times \U_\mathbf{d} \times \dots \textrm{k times}) \rightarrow \U_D$ and is defined as 
\begin{equation}
    \f_k(x,m_1,\dots,m_k) = \R m_k\dots \R m_1 x.
\end{equation}

\subsection*{Existence of an Invariant Measure}
\begin{definition}[Invariant measure \cite{meyn_markov_2009}]
    A Markov chain is said to admit an $\sigma$-invariant measure $\pi$ on $\bX$ if, $\forall A \in \bX$,
\begin{equation}
    \label{eq:invariant_condition}
    \pi (A) = \int_\X \pi(dx) P(x,A).
\end{equation}
\end{definition}
\begin{theorem}
    The Haar measure $\mu$ on $\X = \U_D$ is an invariant measure of the Markov chain in \eqref{eq:markov}.
\end{theorem}
\begin{proof}
    Let $\mu_\mathbf{d}$ denote the Haar measure on $\U_\mathbf{d}$.
    We can write
    \begin{align*}
        \int_\X \pi(dx) P(x,A)  &= \int_{\U_D} \mu(dx) \int_{\U_\mathbf{d}} \mu_\mathbf{d}(dm) \mathbbm{1}(\R mx\in A) \\ 
                                &= \int_{\U_\mathbf{d}} \mu_\mathbf{d}(dm) \int_{\U_D} \mu(dx)  \mathbbm{1}(x\in m^H \R^H A) \\
                                &= \int_{\U_\mathbf{d}} \mu_\mathbf{d}(dm) \mu(m^H \R^H A) \\
                                &= \int_{\U_\mathbf{d}} \mu_\mathbf{d}(dm) \mu(A) \\
                                &= \mu(A) \int_{\U_\mathbf{d}} \mu_\mathbf{d}(dm) =  \mu(A) \mu_\mathbf{d}(\U_\mathbf{d})\\
                                &= \mu(A),
    \end{align*}
    where we have used the invariance property of the Haar measure: for any unitary matrix $V\in\U_D$ and set $B\subseteq \U_D$, $\mu(VB)=\mu(BV)=\mu(B)$; for any unitary matrix $V\in\U_\mathbf{d}$ and set $B\subseteq \U_\mathbf{d}$, $\mu_\mathbf{d}(VB)=\mu_\mathbf{d}(BV)=\mu_\mathbf{d}(B)$.
    \end{proof}

\subsection*{Forward Accessibility, $\psi$-Irreducibility, and Aperiodicity}
Following \cite{meyn_markov_2009}, we define $\Ap^k(x), k\geq1$ as the set of all reachable states from $x$ after $k$ steps:
\begin{equation}
\Ap^k(x) = \{\f_k(x,m_1,\dots,m_k): m_i \in \U_\mathbf{d}, 1\leq i \leq k \},
\end{equation}
$\Ap(x)$ as the set of all reachable states for any number of forward steps:
\begin{equation}
    \Ap(x) = \bigcup_{k = 0}^{\infty} \Ap^k(x).
\end{equation}
We also define $A_{1+}$ as the set of all reachable states of $\mathbf{I}_D$ in one or more steps:
\begin{equation}
    A_{1+} = \bigcup_{k = 1}^{\infty} \Ap^k(\mathbf{I}_D).
\end{equation}
$\Ap(x)$ can be written as ${x}\cup(A_{1+})  x$.

\begin{definition}[Forward accessibility \cite{meyn_markov_2009}]
    A Markov chain is said to be \textit{forward accessible} if, for each $x_0\in \U_D$, $\Ap(x_0)$ has a non-empty interior.
\end{definition}
\begin{definition}[$\psi$-irreduciblity \cite{meyn_markov_2009}]
    A Markov chain is said to be $\phi$-\textit{irreducible} if there exists a measure $\phi$ on $\bX$ such that whenever $\phi(A)>0$, we have $L(x,A)>0$ for all $x\in\X$, where $L(x,A)$ is the return time probability:
    \begin{align*}
        L(x,A)  &=\prob(\tau_A <\infty| \bPhi_0 = x)\\
                &=\prob(\bPhi \textrm{ ever enters }A),
    \end{align*}
    where $\tau_A$ is the first return time, $\tau_A=\min\{n\geq 1: \bPhi_n\in A\}$.
    A Markov chain is $\psi$-irreducible if it is $\phi$-irreducible for some measure $\phi$. The measure $\psi$ is a maximal irreducibility measure.
\end{definition}
For the sake of brevity, the definition of aperiodicity is not repeated here and can be found in \cite{meyn_markov_2009}.

\begin{theorem}
    $\bPhi$ in \eqref{eq:markov} is forward accessible.
\end{theorem}
\begin{proof}
    Lemma \ref{lemma:A1_subgroup} proves that $A_{1+}$ is a subgroup of $\U_D$ that contains the block-diagonal subgroup $\U_\mathbf{d}$.  
    
    $\R\in A_{1+}$ satisfies the strong scrambler design criteria and $\R$ is not a part of any block-diagonal proper subgroup of $\U_D$. 
    Therefore, using the theorems of Borevich and Krupetskii in \cite{borevich_subgroups_1981}, we obtain that $A_{1+}$ is the group $\U_D$ itself.
    $A_{1+} = \U_D \implies \Ap(x) = \U_D,~\forall x\in \U_D$.
    \item We use the following properties of topological groups (CITE):
    \begin{enumerate}
        \item Every subgroup of a topological group is either clopen or has empty interior.
        \item Proper subgroups of a connected topological group have empty interior.
        \item $\U_D$ is a compact, connected topological group.
    \end{enumerate}
    \begin{remark}
        If Postulate \ref{pos:A1_subgroup} holds, then the theorem is proved for any $\R\in \U_D$.
    \end{remark}
\end{proof}

\begin{lemma}
\label{lemma:A1_subgroup}
    $A_{1+}$ is a subgroup of $\U_D$ when the phases of the eigenvalues of $\R$ are rational multiples of $2\pi$.
\end{lemma}
\begin{proof}
        The three group axioms are associativity, existence of identity, and existence of inverse.
        Closure and associativity are proved below and hold for both Lemma \ref{lemma:A1_subgroup} and Postulate \ref{pos:A1_subgroup}
\begin{description}
    \item[Closure]
        $A_{1+}$ is closed under matrix multiplication.
    Consider $a_1,a_2\in A_{1+}$. They can be written as 
    \begin{align*}
        a_1&=\f_{k_1}(\mathbf{I}_D,m^{(1)}_{1},\dots,m^{(1)}_{k_1}),\\
        a_2&=\f_{k_2}(\mathbf{I}_D,m^{(2)}_{1},\dots,m^{(2)}_{k_2}),
    \end{align*}
    for some positive integers $k_1,k_2$, and $m^{(1)}_{1},\dots,m^{(1)}_{k_1}, m^{(2)}_{1},\dots,m^{(2)}_{k_2}\in \U_\mathbf{d}$. 
    Let $k=k_1+k_2$, we have
    \begin{align*}
        &(a_1 a_2) = \f_{k}(\mathbf{I}_D,m^{(1)}_{1},\dots,m^{(1)}_{k_1}, m^{(2)}_{1},\dots,m^{(2)}_{k_2}) \in A_{1+} \\
        &(a_2 a_1) = \f_{k}(\mathbf{I}_D,m^{(2)}_{1},\dots,m^{(2)}_{k_2}, m^{(1)}_{1},\dots,m^{(1)}_{k_1}) \in A_{1+},
    \end{align*}
    thus proving closure.

    \item[Associativity]
        Consider $a_1,a_2,a_3 \in A_{1+}$. They can be written as 
    \begin{align*}
        a_1&=\f_{k_1}(\mathbf{I}_D,m^{(1)}_{1},\dots,m^{(1)}_{k_1}),\\
        a_2&=\f_{k_2}(\mathbf{I}_D,m^{(2)}_{1},\dots,m^{(2)}_{k_2}),\\
        a_3&=\f_{k_3}(\mathbf{I}_D,m^{(3)}_{1},\dots,m^{(3)}_{k_3}),
    \end{align*}
    for some positive integers $k_1,k_2,k_3$, and $m^{(1)}_{1},\dots,m^{(1)}_{k_1}, m^{(2)}_{1},\dots,m^{(2)}_{k_2},m^{(3)}_{1},\dots,m^{(3)}_{k_3} \in \U_\mathbf{d}$.
    Let $k=k_1+k_2+k_3$, we have
    \begin{align*}
        &\f_{k}(\mathbf{I}_D,m^{(1)}_{1},\dots,m^{(1)}_{k_1}, m^{(2)}_{1},\dots,m^{(2)}_{k_2},m^{(3)}_{1},\dots,m^{(3)}_{k_3}) \\
        &\qquad= (a_1 a_2) a_3 \\
        &\qquad= a_1 (a_2 a_3),
    \end{align*}
    thus proving associativity.

    \item[Existence of Identity and Inverse] 
    The phases $\{\theta_i\}$ of the eigenvalues of $\R$ are rational multiples of $2\pi$.
    Therefore, for some finite positive integer $n$,
    \begin{equation*}
        \f_n(\mathbf{I}_D,\mathbf{I}_D,\dots,\mathbf{I}_D) = \R^n = \mathbf{I}_D \in A_{1+},
    \end{equation*}
    thus proving the existence of identity. Moreover, we can see that, for any $m\in \U_\mathbf{d}$,
    \begin{equation*}
        \f_{n}(\mathbf{I}_D,m,\mathbf{I}_D,\dots,\mathbf{I}_D) = \R^{n} m = m.
    \end{equation*}
    Therefore, $\U_\mathbf{d}\subset A_{1+}$.
    Furthermore, $\R^H = \f_{n-1}(\mathbf{I}_D,\mathbf{I}_D,\dots,\mathbf{I}_D) \in A_{1+}$.
    As a result, for any $a=\f_{k}(\mathbf{I}_D,m_{1},\dots,m_{k}) \in A_{1+}$, $a^H$ can be written as
    \begin{equation*}
        a^H = m_1^H \R^H \dots m_k^H \R^H.
    \end{equation*}
    As each of $m_1^H,\dots,m_k^H,\R^H \in A_{1+}$, we have $a^H \in A_{1+}$, thus proving the existence of inverse.
\end{description}
\end{proof}
\begin{postulate}
\label{pos:A1_subgroup}
    $A_{1+}$ is a subgroup of $\U_D$ for any $\R\in \U_D$.
\end{postulate}
\begin{arguments}
    Closure and associativity proved above hold for any $\R\in \U_D$.
    \begin{description}
        \item[Existence of Identity and Inverse] 
        For any $\R\in \U_D$, we postulate that we can find a diagonal unitary matrix $\bm{\Lambda}$ such that $\R \bm{\Lambda}$ has eigenvalues of the form $\{e^{\jmath \theta_i}, 1\leq i \leq D\}$ such that $\{\theta_i\}$ are rational multiples of $2\pi$. 
        For some finite positive integer $n$,
        \begin{equation*}
            \f_n(\mathbf{I}_D,\bm{\Lambda},\dots,\bm{\Lambda}) = \left(\R \bm{\Lambda}\right)^n = \mathbf{I}_D \in A_{1+}.
        \end{equation*}
        Thus proving the existence of identity. Moreover, we can see that, for any $m\in \U_\mathbf{d}$,
        \begin{equation*}
            \f_{n}(\mathbf{I}_D,\bm{\Lambda}m,\bm{\Lambda},\dots,\bm{\Lambda}) = \left(\R \bm{\Lambda}\right)^{n}m = m.
        \end{equation*}
        Therefore, $\U_\mathbf{d}\subset A_{1+}$.
        Furthermore, $\R^H = \bm{\Lambda} \f_{n-1}(\mathbf{I}_D,\bm{\Lambda},\dots,\bm{\Lambda}) \in A_{1+}$.
        As a result, for any $a=\f_{k}(\mathbf{I}_D,m_{1},\dots,m_{k}) \in A_{1+}$, $a^H$ can be written as
        \begin{equation*}
            a^H = m_1^H \R^H \dots m_k^H \R^H.
        \end{equation*}
        As each of $m_1^H,\dots,m_k^H,\R^H \in A_{1+}$, we have $a^H \in A_{1+}$, thus proving the existence of inverse.
    \end{description}
\end{arguments}
\begin{corollary}
    $\bPhi$ in \eqref{eq:markov} is $\psi$-irreducible and aperiodic.
\end{corollary}
\begin{proof}
    The set $A_{1+}$ is a \textit{minimal} set for $\bPhi$, i.e., it is closed, invariant, and does not contain any closed invariant set as a proper subset. $\psi$-Irreducibility and aperiodicity are direct properties of a closed and connected minimal set \cite[Theorem 7.2.4, Proposition 7.3.4]{meyn_markov_2009}. 
\end{proof}

\begin{theorem}
    $\bPhi$ in \eqref{eq:markov} is $\psi$-irreducible for $\psi=\mu$, the Haar measure on $\U_D$.
\end{theorem}
\begin{proof}
From Lemma \ref{lemma:A1_subgroup}, $\bPhi$ is $\phi$-irreducible for $\phi=\mu$. This is because, for any $A\in\mathcal{B}(\U_D)$ such that $\mu(A)>0$, is reachable from every $x\in \U_D$. In other words, $L(x,A)>0$.

The measure $\psi$ is equivalent to \cite{meyn_markov_2009}
\begin{equation}
\label{eq:psi_maximal}
    \psi^{\prime}(A) = \int_{\X}\phi(dy)K_{\frac{1}{2}}(y,A),
\end{equation}
where, for $x\in \X, A\in \mathcal{B}(\U_D)$,
\begin{align}
    K_{\frac{1}{2}}(x,A) &= \sum_{n=0}^{\infty} P^n(x,A) 2^{-(n+1)}.
\end{align}
Substituting $\phi=\mu$ and $\X = \U_D$ in \eqref{eq:psi_maximal} we get,
\begin{align*}
    \psi^{\prime}(A)    &= \int_{\U_D}\mu(dy)K_{\frac{1}{2}}(y,A) \\
                        &= \sum_{n=0}^{\infty} 2^{-(n+1)} \int_{\U_D}\mu(dy) P^n(y,A) \\
                        &= \sum_{n=0}^{\infty} 2^{-(n+1)} \mu(A) \\
                        &= \mu(A).
\end{align*}
    
\end{proof}

\subsection*{Recurrence and Positive Harris Property}
\begin{definition}[Occupation time]
    For any set $A\in\mathcal{B}(\U_D)$, the occupation time $\eta_A$ is the number of visits by $\bPhi$ to $A$ after time step zero:
    \begin{equation}
        \eta_A = \sum_{n=1}^{\infty} \mathbbm{1}(\bPhi_n\in A).
    \end{equation}
\end{definition}
\begin{definition}[Positive chain]
    A Markov chain is said to be \textit{positive} if it is $\psi$-irreducible and admits an invariant probability measure \cite{meyn_markov_2009}.  
\end{definition}
\begin{definition}[Harris]
    A set $A$ is called \textit{Harris recurrent} if it is visited infinitely often
    \begin{align*}
        \prob(\eta_A=\infty|\bPhi_0\in A).
    \end{align*}
    A Markov chain is \textit{Harris} (recurrent) if it is $\psi$-irreducible and every set $A$ such that $\psi(A)>0$ is Harris recurrent \cite{meyn_markov_2009}.
\end{definition}
\begin{theorem}
    $\bPhi$ in \eqref{eq:markov} is positive Harris.
\end{theorem}
\begin{proof}
    Lemma \ref{lemma:feller} proves that $\bPhi$ is a Feller chain. 
    $\bPhi$ is $\psi$-irreducible according to the Haar measure and the support of $\psi$ is $\U_D$ which has a non-empty interior. 
    As a result, $\bPhi$ is a $\psi$-irreducible \textit{T-chain} \cite[Theorem 6.01]{meyn_markov_2009}.
    $\bPhi$ is \textit{non-evanescent} as $\prob(\bPhi\rightarrow\infty|\bPhi_0=x)=0, \forall x\in\U_D$. 
    Non-evanescent $\psi$-irreducible T-chains are Harris (recurrent) \cite[Theorem 9.2.2]{meyn_markov_2009}).
\end{proof}
\begin{definition}[Feller]
    The transition kernel $P$ acts on a bounded function $h$ through the following mapping. For $x\in \X$,
    \begin{equation}
        Ph(x) = \int P(x,dy) h(y).
    \end{equation}
    A Markov chain is said to be \textit{Feller} if $P$ maps the set of bounded continuous functions $\mathcal{C}(\X)$ to $\mathcal{C}(\X)$.
\end{definition}
\begin{lemma}
    \label{lemma:feller}
    $\bPhi$ in \eqref{eq:markov} is a Feller chain.
\end{lemma}
\begin{proof}
    Consider a bounded continuous function $h:\U_D \rightarrow \U_D$.
    \begin{align*}
        Ph(x)   &= \int_{\U_D} P(x,dy) h(y) \\
                &= \int_{\U_D} \int_{\U_\mathbf{d}} \mu_\mathbf{d}(dm) \mathbbm{1}(\R mx\in dy) h(y) \\
                &= \int_{\U_\mathbf{d}} \mu_\mathbf{d}(dm) \int_{\U_D} \mathbbm{1}(\R mx\in dy) h(y) \\
                &= \int_{\U_\mathbf{d}} \mu_\mathbf{d}(dm) h(\R mx) \\
                &= \int_{\U_\mathbf{d}} \mu_\mathbf{d}(dm) (h \circ \f(\bm{\cdot},m))(x) \\
                &= \E{h \circ \f(\bm{\cdot},\mathbf{M})}(x),
    \end{align*}
    where the expectation is over the $\mathbf{d}$-block unitary matrices distributed according to the Haar measure $\mu_\mathbf{d}$. $Ph$ is continuous and bounded as $h$ and $f(\bm{\cdot},m), m\in\U_\mathbf{d}$ are continuous and bounded.
\end{proof}

\subsection*{Uniform Ergodicity}
\begin{remark}
    The aperiodic ergodic theorem \cite[Theorem 13.0.1]{meyn_markov_2009} says that for an aperiodic Harris recurrent chain $\bPhi$ with an invariant probability measure $\pi$ is positive Harris and the invariant measure is unique (up to a scaling) and finite such that for every initial condition $x_0\in \X$, 
    \begin{equation}
        \sup_{A\in\bX} \left| P^n(x_0,A) - \pi(A) \right| \rightarrow 0,
    \end{equation}
    as $n\rightarrow \infty$. The chain is also said to be \textit{ergodic}.
\end{remark}

\begin{definition}[V-uniformly ergodic \cite{meyn_markov_2009}]
    Define the $V$-norm distance between two probability transition functions $P_1$ and $P_2$ and for a positive function $\infty>V\geq 0$ as follows,
    \begin{equation}
        |||P_1 - P_2|||_V = \sup_{x\in \X} \frac{\|P_1(x,\bm{\cdot}) - P_2(x,\bm{\cdot})\|_V}{V(x)},
    \end{equation}
    and the outer product of the function $1$ and the measure $\pi$
    \begin{align}
        [1\otimes \pi](x,A)&=\pi(A), &x\in\X, A\in \bX.
    \end{align}
    An ergodic chain $\bPhi$ is called $V$\textit{-uniformly ergodic} if, for $n\rightarrow \infty$,
    \begin{equation}
        |||P^n - 1\otimes\pi|||_V \rightarrow 0.
    \end{equation}
\end{definition}
\begin{remark}
    Suppose $\bPhi$ is $\psi$ irreducible and aperiodic, then for $V\geq 1$, $\bPhi$ is $V$-uniformly ergodic \cite[Theorem 16.0.1]{meyn_markov_2009}.
\end{remark}

\subsection*{Law of Large Numbers and Central Limit Theorem}
We are interested in a series $S_n(g)$ for a function $g$ on $\X=\U_D$,
\begin{equation}
    S_n(d) = \sum_{k=1}^{n} g(\bPhi_k).
\end{equation}
\begin{definition}[Law of Large Numbers]
    We say that the LLN holds if
    \begin{align}
        &\lim_{n\rightarrow \infty} \frac{1}{n} S_n(g) = \pi(g)\qquad\textrm{a.s.}
    \end{align}
\end{definition}
We now state sufficient conditions for the LLN to hold for a Markov chain \cite[Theorem 17.0.1]{meyn_markov_2009}:
\begin{itemize}
    \item $\bPhi$ is a positive Harris chain.
    \item $\bPhi$ has an invariant probability $\pi$.
    \item $\pi(|g|) < \infty$.
\end{itemize}
\begin{definition}[Central Limit Theorem]
    We say that the CLT holds if there exists a constant $0<\gamma_g^2 <\infty$ such that for each $x\in \X$,
    \begin{equation}
        \frac{1}{\sqrt(n) \gamma_g} S_n(\Bar{g}) \xrightarrow{\textrm{distr.}} \mathcal{N}(0,1),
    \end{equation}
    where $\Bar{g}=g - \pi(g)$.
\end{definition}
We now state sufficient conditions for the CLT to hold for a Markov chain \cite[Theorem 17.0.1]{meyn_markov_2009}:
\begin{itemize}
    \item $\bPhi$ is a positive Harris chain.
    \item $\bPhi$ has an invariant probability $\pi$.
    \item $\bPhi$ is $V$-uniformly ergodic.
\end{itemize}
When these conditions are satisfied, then, for any $g^2\leq V$,
$\gamma_g^2 = \E[\pi]{\Bar{g}^2(\bPhi_0)} + 2\sum_{k=1}^{\infty} \E[\pi]{\Bar{g}(\bPhi_0) \Bar{g}(\bPhi_k)}$
is well-defined, non-negative and finite and coincides with the asymptotic variance:
\begin{equation*}
    \lim_{n\rightarrow \infty} \frac{1}{n} \E[\pi]{S_n^2(\Bar{g})} = \gamma_g^2.
\end{equation*}
The CLT holds when $\gamma_g^2 > 0$. If $\gamma_g^2 = 0$, then 
\begin{align*}
    &\lim_{n\rightarrow \infty} \frac{1}{\sqrt{n}} S_n(g) = 0\qquad\textrm{a.s}.
\end{align*}
The group delay operator $\mathbf{G}_\textrm{tot}$ can be written as a function of $\{\bPhi_n\}$.
For each matrix element $\mathbf{G}_\textrm{tot}[i,j], i,j=1,\dots,D$, a function $g_{i,j}$ can be defined. 
We know from Appendix \ref{app:RMS_GDS} that
\begin{align*}
    &K\rightarrow \infty \implies \E{\mathbf{G}_\textrm{tot}(K)}/K \rightarrow \bm{0}_{D\times D}.
\end{align*}
Therefore, $\Bar{g}_{i,j} = 0$.
Similarly, as
\begin{align*}
    &K\rightarrow \infty \implies \E{\mathbf{G}_\textrm{tot}^H \mathbf{G}_\textrm{tot}(K)}/K \rightarrow \sigma^2_{\mathrm{GD}}(K) \mathbf{I}_D.
\end{align*}
Therefore, $\gamma_{g,(i,j)}^2 = \left(\sum_{l=1}^{N_g} \sigma_{\text{intra},l}^2 + \tautau_0^T \mathbf{D} \tautau_0 \right)/D$.
Combining these observations with the fact that $\mathbf{G}_\textrm{tot}$ is Hermitian symmetric, we can conclude that $\mathbf{G}_\textrm{tot}(K)$ approaches a zero-trace Gaussian unitary ensemble as $K\rightarrow \infty$.

\section{Amplified Spontaneous Emission Noise}
\label{app:ASE}
Following the analysis in \cite[Discussion]{ho_mode-dependent_2011}, we can write the contribution of the $k$th amplifier to the output noise as
\begin{equation}
    \mathbf{M}_K \R \dots \R \mathbf{M}_{k+1} \bm{\Sigma_0}\mathbf{n}_k,
\end{equation}
where $\mathbf{n}_k\sim \mathcal{N}(\bm{0}_{D\times 1}, \mathbf{I}_D)$, and $\bm{\Sigma}_0 = \diag{\sigma_1,\dots,\sigma_D}$ is the diagonal matrix of noise standard deviations of the local uncoupled modes.
The corresponding noise correlation matrix is equal to
\begin{equation}
    \bm{\Sigma}_k = \mathbf{M}_K \R \dots \mathbf{M}_{k+1} \bm{\Sigma}_0^2 \mathbf{M}_{k+1}^H \dots \R^H \mathbf{M}^H_K.
\end{equation}
Similar to the analysis in Appendix \ref{app:RMS_GDS}, taking an expectation over all random realizations of $\{\mathbf{M}_k\}$ yields, for $1\leq k \leq K-1$:
\begin{equation}
    \E{\bm{\Sigma}_k} =  \left(\left(\DPC\right)^{K-k-1} \bm{\kappa}_0\right) \cdot \Vec{\bm{\Gamma}},
\end{equation}
where $\bm{\kappa_0}$ is the $N_g$-dimensional mode-group-averaged noise variances vector:
\begin{equation*}
    \kappa_{0,i}=\frac{1}{d_i}\sum_{l\in \mathcal{M}_i} \sigma^2_l.    
\end{equation*}
The overall expected noise correlation matrix can be written as:
\begin{align}
    \begin{split}
        \sum_{k} \E{\bm{\Sigma}_k} =& \bigg( \bm{\kappa}_0 + \DPC \bm{\kappa}_0 + \dots\\
        & \dots + (\DPC)^{K-2}\bm{\kappa}_0 \bigg) \cdot \Vec{\bm{\Gamma}}.
    \end{split}
\end{align}
In the scenario of strong mode scrambling, for large $K$, $(\DPC)^K$ approaches a rank one matrix. 
The term $(\bm{\kappa}_0 + \DPC \bm{\kappa}_0 + \dots + (\DPC)^{K-2}\bm{\kappa}_0)$ tends to a constant times the vector $K \mathbf{1}_{N_g\times 1}$.
Therefore, $\sum_{k} \E{\bm{\Sigma}_k}$ tends to a constant times $K \mathbf{I}_{D}$.

\end{document}